\documentclass[a4paper,twocolumn,11pt,accepted=2021-08-12]{quantumarticle}
\pdfoutput=1
\usepackage[utf8]{inputenc}
\usepackage[english]{babel}
\usepackage[T1]{fontenc}
\usepackage{amsmath}
\usepackage{amssymb}
\usepackage[bookmarks=false]{hyperref}
\usepackage[numbers,sort&compress]{natbib}

\usepackage{tikz}
\usepackage{lipsum}
\usepackage{xcolor}

\begin{document}

\title{Quadrature protection of squeezed states in a one dimensional photonic topological insulator}

\author{J. Medina Dueñas}
\email{joamedinaduenas@gmail.com}
\author{G. O'Ryan P\'{e}rez}
\affiliation{Departamento de F\'{\i}sica,  Facultad
de Ciencias F\'isicas y Matem\'aticas, Universidad de Chile,
Santiago, Chile}
\author{Carla Hermann-Avigliano}
\affiliation{Departamento de F\'{\i}sica,  Facultad
de Ciencias F\'isicas y Matem\'aticas, Universidad de Chile,
Santiago, Chile}
\affiliation{ANID - Millenium Science Iniciative Program - Millenium Institute for Research in Optics}
\author{L. E. F. Foa Torres}
\affiliation{Departamento de F\'{\i}sica,  Facultad
de Ciencias F\'isicas y Matem\'aticas, Universidad de Chile,
Santiago, Chile}

\maketitle

\begin{abstract}

What is the role of topology in the propagation of quantum light in photonic lattices? We address this question by studying the propagation of squeezed states in a topological one-dimensional waveguide array, benchmarking our results with those for a topologically trivial localized state, and studying their robustness against disorder. Specifically, we study photon statistics, one-mode and two-mode squeezing, and entanglement generation when the localized state is excited with squeezed light. These quantum properties inherit the shape of the localized state but, more interestingly, and unlike in the topologically trivial case, we find that propagation of squeezed light in a topologically protected state robustly preserves the phase of the squeezed quadrature as the system evolves. We show how this latter topological advantage can be harnessed for quantum information protocols.

\end{abstract}

During the last decades, a series of discoveries illuminated a way where topological arguments allowed to explain or envision the behavior of electrons in materials~\cite{von_klitzing_new_1980,thouless_quantized_1982,haldane_model_1988,kane_quantum_2005,c_l_kane_colloquium_2010}. Originally discovered in solid state physics, these properties are actually fundamental to wave phenomena within periodic media, allowing for the extension of topological physics to photonics~\cite{ozawa_topological_2019,xie_photonics_2018,lu_topological_2014}, mechanical systems~\cite{huber_topological_2016}, and circuits~\cite{lee_topolectrical_2018}, among others~\cite{peano_topological_2015,khanikaev_topologically_2015,wang_topological_2018}; where the exploitation of topologically protected boundary states stands as a key aim across these diverse playgrounds. Topological phases have been found in all kinds of crystal and photonic lattices: in one~\cite{atala_direct_2013,malkova_observation_2009}, two~\cite{konig_quantum_2007,hafezi_robust_2011}, and three~\cite{hsieh_topological_2008,lu_symmetry-protected_2016,hasan_discovery_2017,yang_realization_2019} dimensions; exhibiting topological phases protected by preserved~\cite{kane_quantum_2005, hafezi_robust_2011} or broken~\cite{haldane_model_1988,jotzu_experimental_2014,haldane_possible_2008,raghu_analogs_2008,wang_observation_2009} time-reversal invariance, or by crystal symmetries~\cite{fu_topological_2007,lu_symmetry-protected_2016}; among many others~\cite{rudner_band_2020,giustino_2020_2020,rechtsman_photonic_2013,foa_torres_perspective_2019,weimann_topologically_2017}.

Not only serving as a convenient experimental playground, topological photonics may bring new phenomena unique to quantum states of light. Topology has been found to provide robustness in lasers~\cite{st-jean_lasing_2017, zhao_topological_2018} and amplifiers~\cite{peano_topological_2016-1, wanjura_topological_2020}, while coupling between quantum states of light and topological degrees of freedom allows for topological protection of photon statistics and quantum correlation~\cite{blanco-redondo_topological_2020}. We focus on the latter topic. Research in this topic is only recent, nevertheless, the behavior of single- and bi-photon states has been studied in one- and two-dimensional lattices~\cite{kitagawa_observation_2012,chen_observation_2018,barik_topological_2018,tambasco_quantum_2018,mittal_topological_2018,blanco-redondo_topological_2018,rechtsman_topological_2016,mittal_topologically_2016,wang_topologically_2019}. Promising experimental results show quantum interference of topological states~\cite{tambasco_quantum_2018}, and topological protection of quantum properties, such as spectral~\cite{mittal_topological_2018} and spatial~\cite{blanco-redondo_topological_2018} correlation and entanglement~\cite{rechtsman_topological_2016,mittal_topologically_2016,wang_topologically_2019}. Quantum topological photonics is thus positioned as a potential solution for decoherence-free transport of quantum information even at room temperature~\cite{blanco-redondo_topological_2020}.

Research in this area has been focused on Fock-like states, while the interplay between topology and squeezed light~\cite{peano_topological_2016}, which serves as a fundamental building block for continuous variable quantum information~\cite{braunstein_quantum_2005, gottesman_encoding_2001, menicucci_universal_2006}, remains less explored. Photon correlations in squeezed light generate lower than vacuum noise levels in one quadrature of the field at the expense of an increment in the conjugate quadrature. Hence, squeezed light has been central to important breakthroughs in metrology~\cite{caves_quantum-mechanical_1981,aasi_enhanced_2013}. Furthermore, these correlations may present themselves between distinct modes of the field, giving rise to entangled multi-mode squeezed states. In particular, two-mode squeezed light allows for deterministic generation of entanglement~\cite{menicucci_universal_2006, larsen_deterministic_2019}, and in the high squeezing limit correspond to optic EPR states~\cite{braunstein_quantum_2005}, acquiring a central role in quantum information protocols, such as quantum teleportation~\cite{braunstein_teleportation_1998,yukawa_high-fidelity_2008} and quantum computation~\cite{bourassa_blueprint_2021, larsen_fault-tolerant_2021}. However, the correlated nature of these states makes them sensitive to losses and fabrication imperfections in optic systems~\cite{vahlbruch_observation_2008,li_experimental_2020}. It is therefore of high interest to elucidate the effect of a lattices topology on squeezed light, in the search for topological protection of squeezing.

Here we present a thorough report on the propagation of squeezed light in a topological photonic lattice, and the effects of topology on quantum features of light\footnote{After submission of this work, a related study was reported~\cite{ren_topologically_2021}.}. We study the behavior of photon statistics, squeezing and entanglement in a one-dimensional Su-Schrieffer-Heeger (SSH)~\cite{su_solitons_1979,heeger_solitons_1988,malkova_observation_2009} waveguide array when exciting its topologically protected edge state with quadrature squeezed light, employing analytic and numerical techniques. Our focus is to determine the role that the lattices topology plays in this phenomena, establishing an interplay between lattice symmetries and quantum correlations. To this goal we benchmark our results with those of an impurity induced, topologically trivial localized state, and study their response to disorder. A first conclusion that we extract is that quantum properties inherit the localization of the edge state and therefore follow its fate when disorder is introduced. But since mere localization may be generated and preserved under topologically trivial circumstances, one may wonder if any other more remarkable consequence of topology or \textit{topological advantage} exists. Interestingly, we find that the phase of the squeezed quadrature is preserved in the topological state, providing an advantage for propagation of squeezed light in photonic lattices that may not be replicated by other topologically trivial localized states.

The article is organized as follows. In section \ref{section:SSH} we present the SSH system and briefly discuss its topological properties. We then study the system when the edge state is excited with single-mode squeezed light under two different scenarios. In section \ref{subsection:1mA}, we apply the squeezing operator on the localized eigenmode, studying the distribution of squeezing across the lattice; while in section \ref{subsection:1mB}, pursuing a more feasible method of squeezing the edge state, we inject single mode squeezed light to the edge waveguide and study its propagation. During section \ref{section:two-mode} we turn to the behavior of two-mode squeezing in the topological lattice. We start by addressing the relation of two-mode squeezing and entanglement (section \ref{subsection:2mA}), and later study the propagation of two-mode squeezing along the lattices (section \ref{subsection:2mB}). Finally, in section \ref{section:teleportation} we present an application of our results in a quantum teleportation protocol, where the entanglement resource is topologically protected, which serves as a proof of concept of the relevance of topological protection of quantum information upon practical implementations. The article ends with final remarks and perspectives on section \ref{section:conclusions}.

\section{The SSH lattice}
\label{section:SSH}

The SSH system consists of a one dimensional dimerized chain of identical modes, with a uniform onsite term $\beta$ across the entire lattice and alternating hopping amplitudes $u$ and $v$, described by the hamiltonian
\begin{equation} 
\begin{split}
\mathcal{H}_\text{top} = &\sum_n \beta a_{2n}^\dag a_{2n} + \beta a_{2n+1}^\dag a_{2n+1} + \\ &\left( u a_{2n}^\dag a_{2n+1} + v a_{2n+1}^\dag a_{2n+2} + \text{h.c.} \right) \text{ ,}
\end{split}
\end{equation}
where $a_n$ is the annihilation operator at mode $n$. The system is formed by two sublattices, even and odd modes, with a unit cell containing sites $a_{2n}$ and $a_{2n+1}$. For a waveguide array, the operators $a_n$ are bosonic as each mode represents a waveguide, and $\beta$ corresponds to the propagation constant of the bare waveguides. We take $u$ and $v$ of the same sign. This choice is physically motivated as the hoppings are typically determined by the overlap of the evanescent tails of the modes located at each waveguide~\cite{haus_coupled-mode_1991}. The system presents a bulk bandgap of $|u-v|$, however, exponentially localized states with a propagation constant equal to that of the bare waveguides and perfect sublattice polarization appear at the edge of a finite lattice terminated in a weak coupling. These states are topologically protected by chiral symmetry~\cite{asboth_short_2016}, which is broken by onsite terms differing on each sublattice or couplings between sites of the same sublattice for example. Therefore, introducing disorder to the nearest neighbor hoppings preserves the propagation constant of the edge state, and its sublattice polarization. On the other hand, for symmetry breaking onsite disorder, the topological phase of the lattice is destroyed and these properties fluctuate. We note as well that perfect sublattice polarization can only arise for a semi-infinite lattice as in the finite case there is always a small coupling between states at opposite edges.

\begin{figure}[t]
    \centering
    \includegraphics[width=\columnwidth]{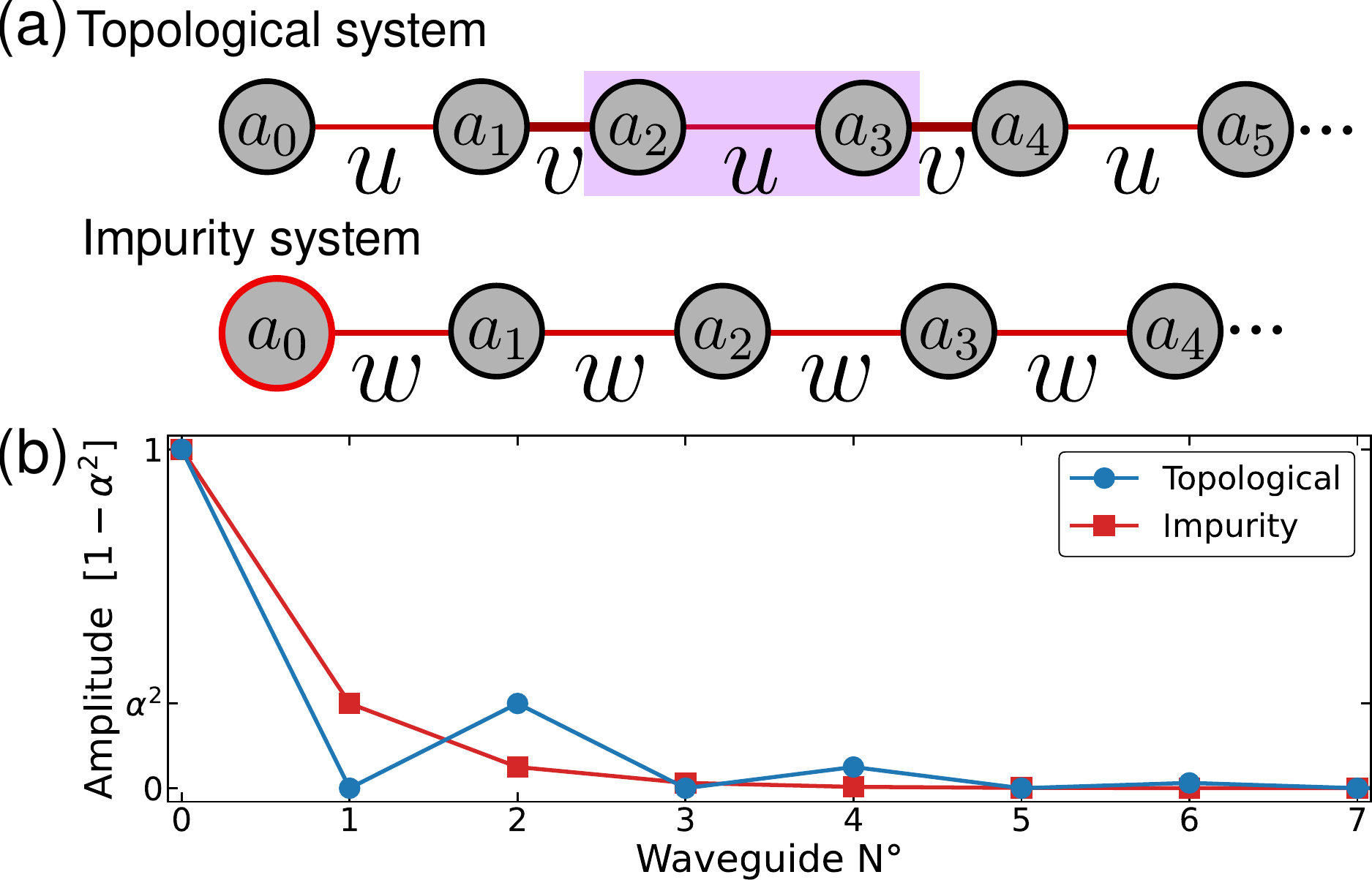}
    \caption{Topological and impurity systems (without disorder) considered through our work. The latter is used to benchmark our results for the topological system. (a) Scheme of the lattices. Sites are represented by grey circles and hoppings by the connecting lines. Unit cell is highlighted in purple. (b) Localized eigenstates probability distribution hosted by each lattice. The states show the same overall exponential decay when adjusting for the different lattice constants, and differ by the sublattice polarization of the topological state.}
    \label{fig:sshlattice}
\end{figure}

For a semi-infinite SSH lattice as depicted in Fig.~\ref{fig:sshlattice}-(a), with $|u|<|v|$, the topological edge state has annihilation operator 
\begin{equation}
A_\text{top} = \sqrt{1 - \alpha^2} \sum_{n \geq 0} (-\alpha)^n a_{2n} \text{ ,}
\label{eq:topologicalstate}
\end{equation}
where $\alpha = u/v$, and the absence of odd modes in $A_\text{top}$ reflects sublattice polarization. The amplitude of the topological state at each waveguide is shown in Fig.~\ref{fig:sshlattice}-(b).

To benchmark our results we need a reference system. Since we are interested in isolating the effect of topology, a natural candidate is a system hosting a topologically trivial edge state. This motivates us to also consider a non-dimerized one-dimensional chain of identical modes, with an onsite impurity $\epsilon_0$, which induces an exponentially localized mode with no topological protection. The hamiltonian of the semi-infinte topologically trivial lattice, with the impurity located at its edge, is
\begin{equation}
\begin{split}
\mathcal{H}_\text{impurity} =\, &\epsilon_0 a_0^\dag a_0\, + \\ &\sum_{n \geq 0} \beta a_n^\dag a_n + \left( w a_{n}^\dag a_{n+1} +  \text{h.c.} \right) \text{ ,}
\end{split}
\end{equation}
with $w$ the hopping amplitude between neighboring waveguides. The localized state of this system is given by its annihilation operator
\begin{equation}
\begin{split}
A_\text{impurity} &= \sqrt{1 - (w / \epsilon_0)^2} \sum_{n \geq 0} (w / \epsilon_0)^n a_{n} \\
&= \sqrt{1 - \alpha^2} \sum_{n \geq 0} (-\alpha)^n a_{n} \text{ ,}
\label{eq:impuritystate}
\end{split}
\end{equation}
where we have chosen $\epsilon_0 = -w / \alpha$ so both localized states present the same spatial distribution, with expception of sublattice polarization. We also set $w = v\alpha / (1 - \alpha)$ so both states present the same spectral separation to the bulk bands. The semi-infinite topologically trivial lattice is depicted in Fig.~\ref{fig:sshlattice}-(a), and the amplitude of its edge state in each waveguide is shown in Fig.~\ref{fig:sshlattice}-(b). In contrast to the topological edge state, the impurity induced one does not present any symmetry protected properties, so its propagation constant and spatial distribution will vary upon any type of disorder present in the lattice.

\section{Single-mode squeezing in a topological state}
\label{section:single-mode}

We now study the behavior of single-mode squeezed light coupled to the topological state under two perspectives: In section \ref{subsection:1mA} we consider the state of the system $|\psi\rangle$ as single-mode squeezed vacuum of the eigenmode, that is $|\psi\rangle = S_{A_{top}}(\xi) |0\rangle$, with $S_a(\xi)$ the single-mode squeezing operator of mode $a$, given by
\begin{equation}
    S_a(\xi) = \exp\left[\frac{1}{2}\left( \xi^* a^2 - \xi (a^\dag)^2 \right)\right] \text{ ,}
    \label{eq:singlemodesqueezingop}
\end{equation}
with $\xi$ the squeezing parameter. Experimentally it would be very challenging to squeeze a collective mode of the lattice; nevertheless, this scenario provides insight on how quantum properties present themselves when exciting solely the topological state. In section \ref{subsection:1mB} we turn to a more feasible scenario, where we excite only the edge waveguide with single-mode squeezed light, and study its evolution.

In both cases, we study the distribution of one- and two-mode squeezing across the lattice, and their response to disorder, comparing with the topologically trivial state. We calculate one-mode squeezing in $dB$ at waveguide $n$ following
\begin{equation}
    S_n = \min_{\phi} \left[ 10 \log_{10} \left( \frac{\langle (\Delta X_n(\phi))^2 \rangle}{\langle(\Delta X)^2\rangle_\text{vacuum}} \right) \right] \text{ ,}
    \label{eq:squeezingdefinition}
\end{equation}
where $\langle(\Delta X)^2\rangle_\text{vacuum} = 1/4$ is the variance of the quadratures for a vacuum state, and $\langle \left(\Delta X_n(\phi) \right)^2 \rangle = \langle X_n(\phi)^2 \rangle - \langle X_n(\phi) \rangle ^2$ is the variance of the field in the $\phi$ rotated quadrature at mode $n$, defined by 
\begin{equation}
    X_n(\phi) = \frac{1}{2}\left[ e^{-i\phi} a_n + e^{i\phi} a_n^\dag \right] \text{ .}
\end{equation}
Non-zero squeezing is obtained when the field presents fluctuations lower than those of vacuum on any quadrature. 

Two-mode squeezing is observed in the superposition of quadratures of both modes. We thus define the two-mode $\phi$ rotated quadrature between waveguides $n$ and $m$~\cite{marino_experimental_2006, rojas-rojas_manipulation_2019}
\begin{equation}
    X_{n, m}(\phi) = \frac{1}{\sqrt{2}}\left[ X_n(\phi) + X_m(\phi) \right] \text{ ,}
\end{equation}
for which two-mode squeezing is calculated in an analogous form to Eq.~\eqref{eq:squeezingdefinition}.

When introducing disorder, we construct a lattice with hopping (onsite) disorder $d$ by sampling each hopping (onsite) term from a uniform distribution of width $d$, centered at the pristine value of the parameter. The results presented for disordered systems correspond to the average over 50 mean values, each of them obtained after averaging the results for six different (a sextet) random realizations of the disordered systems.

\subsection{Squeezing the eigenstate}
\label{subsection:1mA}

We consider semi-infinite lattices, and express their edge states as a linear combination of the operators at each waveguide, $A = \sum c_n a_n$. For the pristine systems the coefficients $c_n$ are given by Eq.~\eqref{eq:topologicalstate} or \eqref{eq:impuritystate} depending on the situation, while for disordered systems they are obtained numerically. For the state $|\psi\rangle =  S_A(\xi) |0\rangle$ we find the following expressions for one- and two-mode squeezing:
\begin{align}
    S_n &= 10 \log_{10}\left( 1 - 2 |c_n|^2 e^{-|\xi|} \sinh|\xi| \right) \text{ ,}
    \label{eq:1msqueezing} \\
    S_{n, m} &= 10\log_{10}\left( 1 - |c_n + c_m|^2 e^{-|\xi|} \sinh|\xi| \right) \text{ .}
    \label{eq:2msqueezing}
\end{align}

\begin{figure}[b!]
    \centering
    \includegraphics[width=\columnwidth]{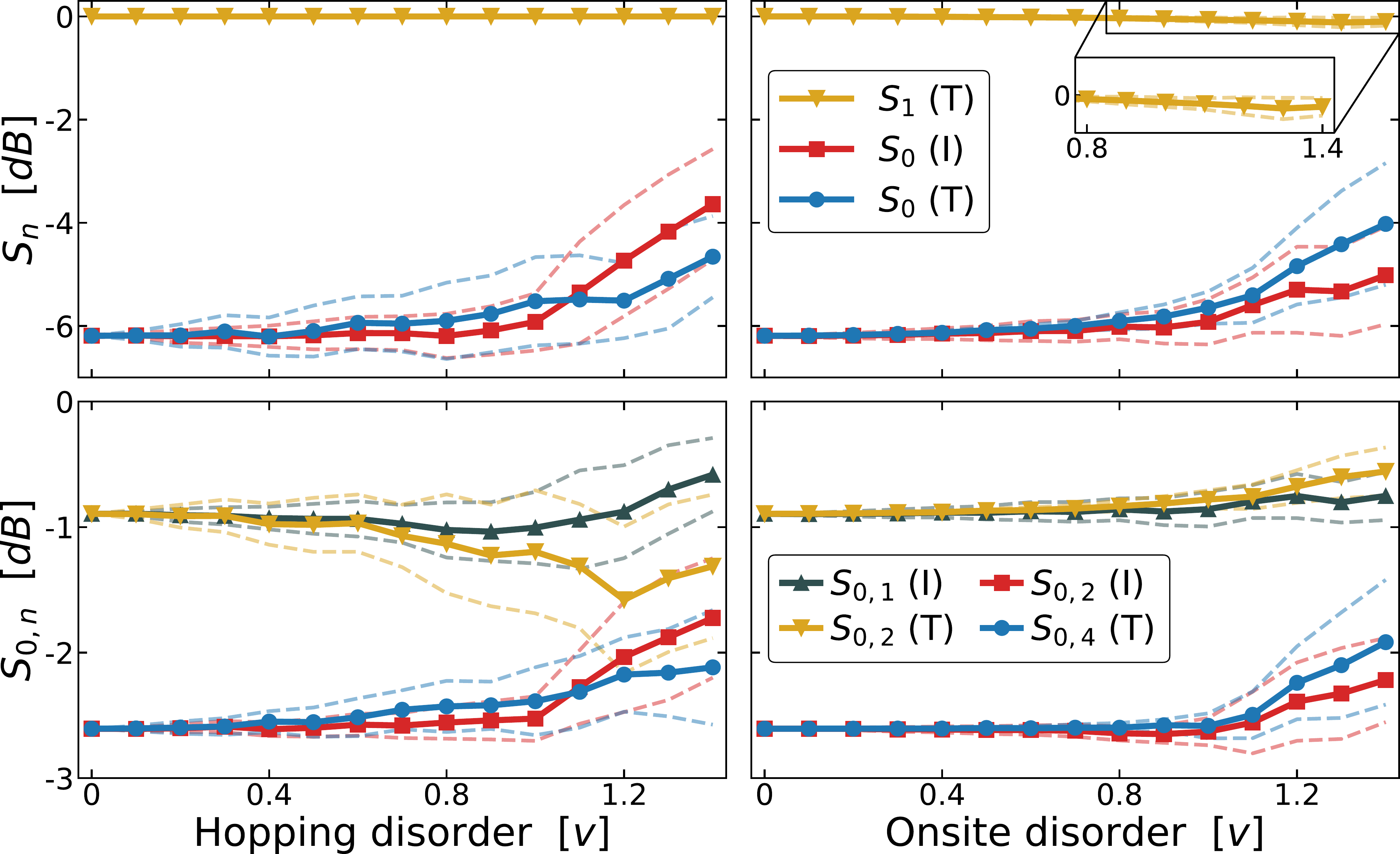}
    \caption{Distribution of squeezing when squeezing the localized eigenstates, for $\alpha=0.3$ and $\xi=0.9$. Solid lines with markers show the statistical average value over disordered systems, and dashed lines represent the confidence interval associated with the standard deviation. Top (bottom) row shows one-mode (two-mode) squeezing. The waveguides at which the observations are made are shown in the legend, with T (I) indicating topological (impurity) state. Zero squeezing at $S_1$(T) reflects sublattice polarization of the topological state, which is broken when introducing onsite disorder as shown in the top right panel. The bottom panels show that two-mode squeezing is sensitive to the phase relation between the modes, obtaining higher squeezing between waveguides 0 and 4 (2) for the topological (impurity) state.}
    \label{fig:ideal}
\end{figure}

Our results in Eq.~\eqref{eq:1msqueezing} and \eqref{eq:2msqueezing}, shown in Fig.~\ref{fig:ideal} as disorder is introduced, indicate that squeezing decays when distancing from the edge, following the exponential decay of the eigenstates. For the pristine lattices, because the topological state at even waveguides shows the same spatial distribution than the impurity induced state, the measured quantities are identical in both systems, but the values obtained at site $2n$ of the topological lattice are obtained at site $n$ of the impurity one. Furthermore, Eq.~\eqref{eq:2msqueezing} reveals that two mode squeezing is sensitive to the phase relation between the individual modes, generating higher squeezing between in-phase modes, being $S_{0, 4}$ ($S_{0, 2}$) the highest, and $S_{0, 2}$ ($S_{0, 1}$) the lowest for the topological (impurity induced) state, as shown in Fig.~\ref{fig:ideal}. For the topological lattice, the absence of one-mode squeezing at odd waveguides reflects sublattice polarization of the state, which is broken when introducing onsite disorder, as occurs in Fig.~\ref{fig:ideal}, top right panel. We observe as well that for high disorder values the states partially delocalize, which is reflected in a decreasing magnitude of one-mode squeezing, while they also show larger statistical fluctuations, represented by wider confidence intervals. Until now, the propagation constant does not seem to take part in the behavior of the systems. In order to observe its effect, we must study the propagation of the states.

\subsection{Propagation of single-mode squeezing}
\label{subsection:1mB}

Since squeezing a collective mode of the system might be experimentally very challenging, we now consider a more feasible state of the field, generated by exciting only the edge waveguide with single mode squeezed vacuum, that is $|\psi(z=0)\rangle = S_{a_0}(\xi) |0\rangle$, and study its evolution along the propagation axis $z$ (see appendix \ref{appendix:A} for numerical methods). We work in a rotated frame, where the maximally squeezed quadrature of the edge waveguide of the pristine lattices corresponds to $X(\phi=0)$ along the entire propagation. We report mean photon number at each quadrature, squeezing at the orthogonal quadratures $X_1=X(0)$ and $X_2=X(\pi/2)$, and maximal squeezing at any rotated quadrature. To maintain this scenario as simple as possible, we consider finite lattices of 15 waveguides each.

\begin{figure}[t]
    \centering
    \includegraphics[width=\columnwidth]{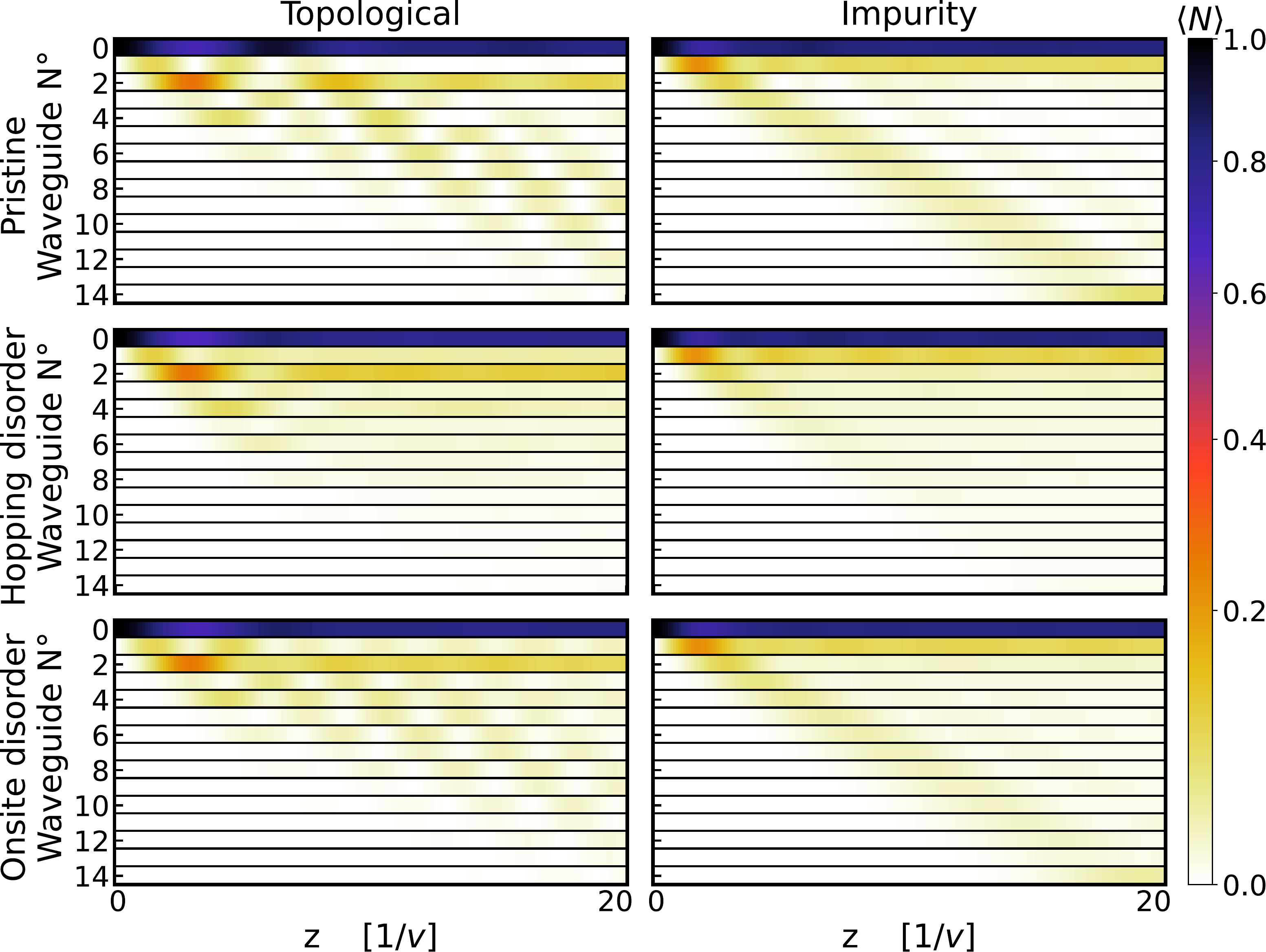}
    \caption{Photon number $\langle N\rangle$ for $\alpha=0.3$ and $\xi=0.9$, in units relative to the initial single-mode squeezed state. Top, middle and bottom row correspond to pristine, hopping disordered ($d=0.6v$) and onsite disordered ($d=0.6v$) lattices respectively, while left and right columns correspond to the topological and impurity systems respectively. Part of the excitation couples to the edge state and remains localized, reflected by a persistent high photon number in the zero-th waveguide. The other part of the excitation transports across the pristine lattices towards the opposite border. Disorder induced Anderson localization obstructs transport across the bulk of the lattices.}
    \label{fig:1m_number}
\end{figure}

The propagation of the excitation across the pristine lattices is shown in Fig.~\ref{fig:1m_number}, top row, depicting the mean photon number in each waveguide as both systems evolve. Part of the excitation remains permanently localized near the edge due to coupling with the localized states, generating a distribution of squeezing qualitatively equal to the one presented in section \ref{subsection:1mA}: Photon number and one-mode squeezing decay while distancing from the edge, respecting sublattice polarization in the topological lattice (see also Fig.~\ref{fig:1m_1msqueezing} in appendix~\ref{appendix:B}). Maximum two-mode squeezing is obtained between waveguides $0$ and $4$ ($2$) in the topological (impurity) state (see Fig.~\ref{fig:1m_2msqueezing} in appendix \ref{appendix:B}). These results show that when injecting one-mode squeezing to the edge waveguide, not only does the excitation remain localized, but the quantum properties of the field inherit the shape of the edge state. This is particularly clear when analyzing two-mode squeezing, which proves to be sensitive to the phase relation between the individual modes. We also observe that part of the excitation does not couple to the edge state, but transports across the lattice, which is explained by the non-perfect overlap between the edge state and the zero-th waveguide.

When introducing disorder we do not observe a significant change in the mean photon number of the localized portion of the excitation, which is coupled to the edge state, for neither type of disorder. On the other hand, as shown in Fig.~\ref{fig:1m_number}, transport of the decoupled excitation is degraded due to Anderson localization of the extended states. In the topological lattice, this causes a persistent non-zero photon number at odd waveguides, even for symmetry preserving hopping disorder.

\begin{figure*}[t]
    \centering
    \includegraphics[width=\textwidth]{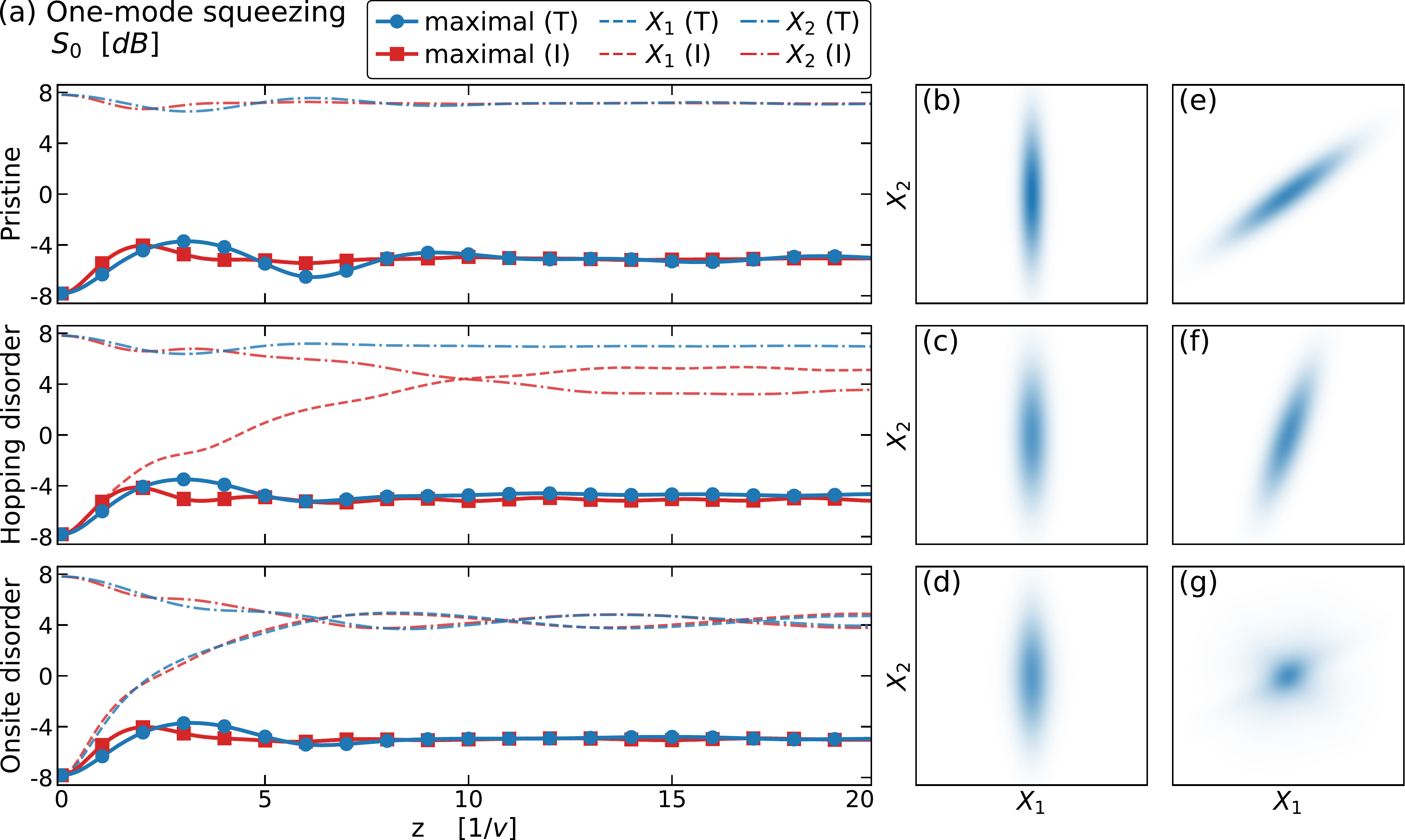}
    \caption{Single-mode squeezing at waveguide zero, for $\alpha=0.3$ and $\xi=0.9$. (a) Solid blue (red) curves with circular (square) marker show squeezing at the maximally squeezed quadrature for the topological (impurity) state. Dashed (dash-dotted) curves show squeezing at quadrature $X_1$ ($X_2$), following the same color assignment as maximal squeezing. The maximally squeezed quadrature of the pristine systems is always $X_1$, as the dashed curves overlap with the solid ones. The maximally squeezed quadrature of the topological state remains $X_1$ as long as chiral symmetry is preserved (hopping disorder), but rotates when it is broken (onsite disorder). The maximally squeezed quadrature of the impurity state rotates upon any form of disorder. Fluctuations in the rotation angle of the squeezed quadrature results in no average squeezing at $X_1$ and $X_2$. Panels (b) to (g) show the Wigner function of the system at the zero-th waveguide (See movie 1 for the evolution of the average Wigner functions). (b) Input single-mode squeezed state, squeezed in $X_1$. (c) One realization of the hopping disordered topological system at $z=10v^{-1}$, which remains squeezed in $X_1$. (d) Average Wigner function for the hopping disordered topological state at $z=10v^{-1}$, showing non-zero squeezing as all realizations remain squeezed in $X_1$ due to topological protection. (e) and (f) Two realizations of the hopping disordered impurity system at $z=10v^{-1}$, which are squeezed at different rotated quadratures. (g) Average Wigner function for the hopping disordered impurity state at $z=10v^{-1}$, showing zero average squeezing as the squeezed quadrature differs between different random realizations.}
    \label{fig:1m_squeezing}
\end{figure*}

The most remarkable consequences of topological protection, or lack of it, are observed when analyzing one-mode squeezing at the edge waveguide, as shown in Fig.~\ref{fig:1m_squeezing}-(a). 
We observe that maximal squeezing in the pristine lattices, which is always obtained in quadrature $X_1$, has an approximately equal value in the topological and impurity state. 
The role of topology in the propagation of squeezed light is revealed when introducing hopping disorder. Maximal squeezing appears to be insensitive to disorder, however, a drastically different behavior is observed for squeezing at quadratures $X_1$ and $X_2$. Because hopping disorder preserves chiral symmetry, the propagation constant of the topological state remains equal to that of the bare waveguides, therefore its maximally squeezed quadrature continues to be $X_1$ along the entire propagation (Fig.~\ref{fig:1m_squeezing}-(b)--(d)). In contrast, random disorder generates fluctuations in the propagation constant of the impurity state, rotating the maximally squeezed quadrature. Even though the behavior of maximal squeezing remains similar to that of the pristine lattice, the rotation angle of the squeezed quadrature differs between different random realizations, thus when averaging them at a fixed quadrature, coherence of the quantum state is lost, and no squeezing is obtained (Fig.~\ref{fig:1m_squeezing}-(e)--(g)).

This discussion is consistent with the results for the onsite disordered lattices. Just as occurred for hopping disorder, the behavior of maximal squeezing is similar to the pristine case. Nevertheless, introduction of onsite disorder breaks chiral symmetry in the SSH lattice, therefore its edge state looses topological protection, and its propagation constant ceases to be topologically locked. The fluctuations in the propagation constant between different random realizations causes different rotation angles of their maximally squeezed quadratures, obtaining no squeezing when averaging over them at a fixed quadrature. These results show that when propagating squeezed light through a localized state in a photonic lattice, topological protection of the state allows for phase coherence in the squeezed quadrature.

\section{Two-mode squeezing of topological states}
\label{section:two-mode}

Having elucidated the role of topology in the propagation of single-mode squeezed light, we now study two-mode squeezing coupled to respective localized states. In contrast to single-mode squeezing, now the main quantum correlations occur between different modes, giving rise to entanglement. As seen in the previous section, the main consequence of the lattices topology is protection of the squeezed quadrature. We therefore begin this section by discussing the effects of rotating the squeezed quadrature on the generation of two-mode squeezing and entanglement for a two-mode squeezed vacuum state (section \ref{subsection:2mA}). We then study the propagation of two-mode squeezing across a topological lattice, when the edge waveguides hosting the localized states are initially squeezed (section \ref{subsection:2mB}).

\subsection{Two-mode squeezing and entanglement}
\label{subsection:2mA}

In this section we introduce some standard definitions. Although this can be found in Refs.~\cite{braunstein_quantum_2005, braunstein_teleportation_1998, simon_peres-horodecki_2000}, we briefly introduce them here to make the discussion self-contained and prepare the reader for the interpretation of our results showed in section \ref{subsection:2mB}.
 
The two-mode squeezed vacuum state between modes $a$ and $b$ of the field is defined by $|\xi_{a,b}\rangle = S_{a, b}(\xi) |0\rangle$, with $S_{a, b}$ the two-mode squeezing operator
\begin{equation}
    S_{a, b}(\xi) = \exp\left[ \xi^* a b - \xi a^\dag b^\dag \right] \text{ .}
\end{equation}
For such a state no one-mode squeezing is measured in the individual modes, as $\langle (\Delta X_\mu(\phi))^2\rangle\geq\langle (\Delta X)^2\rangle_\text{vacuum}$, with $\mu=a,b$. On the other hand, the variance of the two-mode 
quadrature is $\langle(\Delta X_{a,b}(\phi))^2\rangle = \left[ \cosh2|\xi| - \sinh2|\xi| \cos(\theta - 2\phi) \right] / 4$, with $\xi = |\xi|e^{i\theta}$, thus maximal squeezing is measured at $\phi = \theta / 2$.

We quantify entanglement employing the Peres-Horodecki criterion for continuous variables~\cite{simon_peres-horodecki_2000,braunstein_quantum_2005}, stating that the degree of non-separability is measured by the negativity of the partially transposed density matrix. Implementation of this criterion is as follows: We begin defining the $4\times4$ correlation matrix $V$ by $V_{i,j} = \langle \hat\chi_i \hat\chi_j + \hat\chi_j \hat\chi_i \rangle / 2$, with $\hat\chi = (X_a(0), X_a(\pi/2), X_b(0), X_b(\pi/2))$ the basis of the phase-space representation of the system. The partially transposed correlation matrix is obtained by $\tilde{V} = \Gamma V \Gamma$, with $\Gamma = \text{diag}(1, 1, 1, -1)$, which acts by flipping the sign of $X_b(\pi/2)$. Finally, negativity of the partially transposed density matrix is determined by $\tilde{V} - \Lambda \ngeq 0$, with $\Lambda_{i,j} = \langle \hat\chi_i \hat\chi_j - \hat\chi_j \hat\chi_i \rangle / 2$. Therefore, the minimum eigenvalue of $\tilde{V} - \Lambda$ serves as an entanglement measure, revealing non-separability when it takes negative values. 
For two-mode squeezed vacuum, the correlation matrix takes the form $V = \mathcal{R}(\theta) V_0 \mathcal{R}^\dag(\theta)$, with
\begin{widetext}
\begin{equation}
    V_0 = \frac{1}{4} \begin{pmatrix} \cosh2|\xi| & 0 & -\sinh2|\xi| & 0 \\ 0 & \cosh2|\xi| & 0 & \sinh2|\xi| \\ -\sinh2|\xi| & 0 & \cosh2|\xi| & 0 \\ 0 & \sinh2|\xi| & 0 & \cosh2|\xi| \end{pmatrix} \text{ ,}
\end{equation}
\end{widetext}
and $\mathcal{R}(\theta) = \text{diag}(R(\theta/2), R(\theta/2))$, with $R$ the $2\times2$ rotation matrix. We observe that, in agreement with the squeezing measurement, any two-mode squeezed vacuum state can be represented as a $\theta = 0$ state upon a rotation of the phase space basis in $\theta/2$. Logically, the degree of entanglement between the modes does not depend on the basis choice of the system, but, as we will see next, the entangled variables do depend on this.

The Wigner function of the state is a multivariate gaussian distribution
\begin{equation}
    W_{a,b}(\chi) = \frac{|V|^{-1/2}}{(2\pi)^2} \exp\left[ -\frac{1}{2} \chi^T V^{-1} \chi \right] \text{ ,}
\end{equation}
with $\chi = (q_a, p_a, q_b, p_b)$ the corresponding eigenvalues of $\hat\chi$. For $\theta=0$
\begin{equation}
    \begin{split}
    W_{a,b}(\chi) &= \frac{4}{\pi^2} \exp \left\{ -e^{2|\xi|} \left[ (q_a + q_b)^2 + (p_a - p_b)^2 \right] \right. \\ &\left.- e^{-2|\xi|} \left[ (q_a - q_b)^2 + (p_a + p_b)^2 \right] \right\}  \text{ .}
    \end{split}
\label{eq:wigner_2msq}
\end{equation}
Taking the infinite squeezing limit $|\xi| \rightarrow \infty$, which is unphysical as such a state would require infinite energy, but represents a perfectly correlated maximally entangled state, the Wigner function approaches $W \propto \delta(q_a + q_b) \delta(p_a - p_b)$. From this we conclude that the entangled variables of the system are $q_a$ and $q_b$; and $p_a$ and $p_b$. 

If $\theta \neq 0$, the expression for the Wigner function of the state would be the same as in Eq.~\eqref{eq:wigner_2msq}, but for a rotated phase space basis, that is $q_\mu \rightarrow q_\mu \cos(\theta/2) + p_\mu \sin(\theta/2)$, and $p_\mu \rightarrow p_\mu \cos(\theta/2) - q_\mu \sin(\theta/2)$. Thus, the entangled variables correspond to a rotation of the basis, $\mathcal{R}(\theta) \chi$. From this analysis we conclude that, for a two-mode squeezed vacuum state, rotation of the squeezed quadrature does not affect the magnitude of entanglement in the system, but does change the entangled variables. With this in mind, we now study the propagation of two-mode squeezed light in the topological SSH lattice.

\subsection{Propagation of two-mode squeezing}
\label{subsection:2mB}

We consider two independent SSH lattices of 15 waveguides each, which we label as lattices A and B, each one hosting a topologically protected edge state, and study the propagation of two-mode squeezing along the system. For comparison, we also consider topologically trivial lattices hosting impurity induced edge states. The initial state of the field is given by the two-mode squeezed vacuum state between the edge waveguide of both lattices, that is $|\psi(z=0)\rangle = S_{a_0, b_0}(\xi) |0\rangle$. We explore the response of the systems to hopping and onsite disorder, while reporting two-mode squeezing and entanglement between both edge waveguides.

\begin{figure}[t]
    \centering
    \includegraphics[width=\columnwidth]{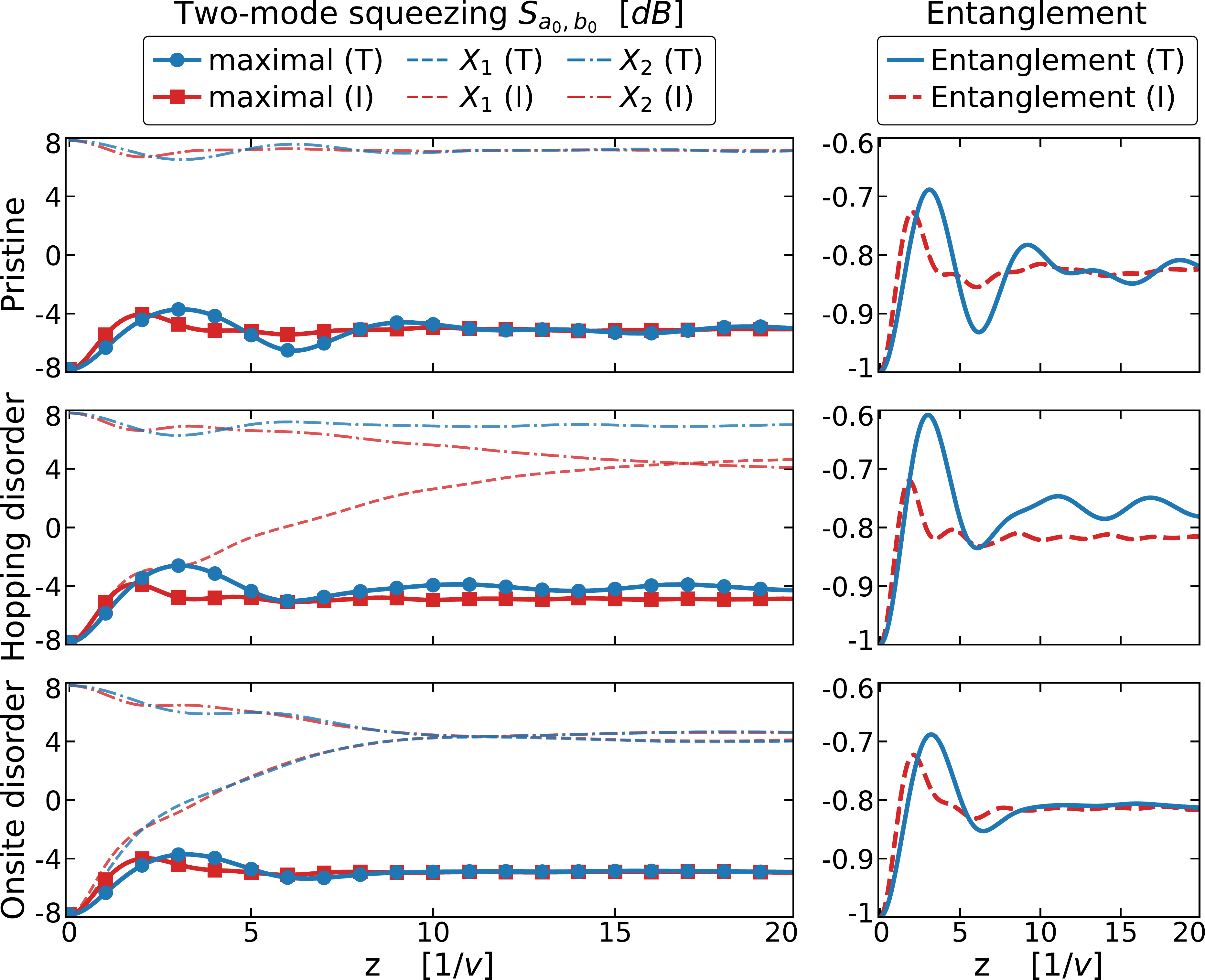}
    \caption{Two-mode squeezing and entanglement between both edge waveguides, for $\alpha=0.3$ and $\xi=0.9$. Left column: solid blue (red) curves with circular (square) marker show squeezing at the maximally squeezed quadrature for the topological (impurity) state. Dashed (dash-dotted) curves show squeezing at quadrature $X_1$ ($X_2$), following the same color assignment as maximal squeezing. Right column: solid blue (dashed red) curves show entanglement for the topological (impurity) state, in units relative to the initial two-mode squeezed vacuum state. Similar to maximal squeezing, the magnitude of entanglement is not sensitive to the topology of the edge state nor the type of disorder, but squeezing at quadratures $X_1$ and $X_2$ reveals topological protection of the maximally entangled variables when chiral symmetry is preserved.}
     \label{fig:2m_squeezing}
\end{figure}

Fig. \ref{fig:2m_squeezing}, left column, shows the evolution of two-mode squeezing along the lattices, whose results agree with those of section \ref{section:single-mode}: Maximal squeezing measured in both systems has an approximately equal value along the propagation direction, which is neither sensitive to topology nor disorder. For the topological system, the maximally squeezed quadrature is $X_1$ as long as chiral symmetry is preserved, because the propagation constant of the edge state is topologically locked. In contrast, if the symmetry is broken by onsite disorder, squeezing measured at quadratures $X_1$ and $X_2$ averages to zero over the random realizations of the system due to fluctuations in the rotation angle of the maximally squeezed quadrature. On the other hand, as the impurity induced edge state does not present any preserved properties, the squeezing values at fixed quadratures are highly sensitive to both types of disorder.

The behavior of entanglement in the systems, portrayed in Fig.~\ref{fig:2m_squeezing}, right column, is qualitatively similar to that of maximal two-mode squeezing: it takes an approximately constant value along the propagation axis, and does not appear to be sensitive to disorder. From this perspective, the results indicate that the total amount of quantum correlations present in the system are independent of the topology of the involved edge states. However, they must be interpreted alongside the information provided by squeezing at the non-rotated quadratures, and the discussion of section \ref{subsection:2mA}. Despite the fact that the amount of entanglement is independent of the topology of the edge state, its topological protection allows to control the maximally entangled variables, protecting them from any symmetry preserving disorder.

\section{Quantum teleportation employing a topological lattice}
\label{section:teleportation}

In this section we demonstrate the previously highlighted consequences of the topological protection of squeezing. For this we implement quantum teleportation of a single-photon, where the two-mode squeezed states studied in section \ref{subsection:2mB} serve as the shared entanglement resource. As we will show later in this section, carrying out the teleportation protocol in a topological lattice turns out to be advantageous compared with a topologically trivial lattice bearing a localized state. This topological advantage is quantified by a robust fidelity in the presence of disorder. In the following we describe the teleportation protocol and expose our main results.

The teleportation protocol has been developed in Ref.~\cite{braunstein_teleportation_1998}. To make the discussion self-contained, in this paragraph we outline the main steps. We wish to to transfer an arbitrary quantum state $|\psi_\text{in}\rangle$ from Alice's location to Bob's, for which we rely on a shared entanglement resource between both parties, given by the state $|\psi_{a,b}\rangle$. We must consider three modes of the field: $a$ and $b$ corresponding to the entangled modes sent to Alice and Bob respectively, and mode $c$ containing the input state. Defining $\chi_\mu = (q_\mu, p_\mu)$ (with $\mu = a, b, c$) the phase space basis of the respective modes, the Wigner function of the system in this initial setup is
\begin{equation}
    W_0\left(\chi_a, \chi_b, \chi_c\right) = W_{a,b}\left(\chi_a, \chi_b \right) W_\text{in}\left(\chi_c \right) \text{ ,}
\end{equation}
where $W_{a,b}$ and $W_\text{in}$ represent the Wigner function of $|\psi_{a,b}\rangle$ and $|\psi_\text{in}\rangle$ respectively. Alice couples her entangled mode with the input state at a 50:50 beam splitter, obtaining the Wigner function
\begin{equation}
    W_1\left(\chi_a, \chi_b, \chi_c\right) = W_0\left( \frac{\chi_a - \chi_c}{\sqrt{2}}, \chi_b, \frac{\chi_c + \chi_a}{\sqrt{2}} \right) \text{ .}
    \label{eq:wigner_1}
\end{equation}
Alice then measures the quadratures corresponding to $q_a$ and $p_c$, which become classically determined random variables, following the probability distribution
\begin{equation}
    \mathcal{P}(q_a, p_c) = \int dp_a\, dq_b\, dp_b\, dq_c\, W_1\left( \chi_a, \chi_b, \chi_c \right) \text{ .}
\end{equation}
Given a pair of measured values $q_a$ and $p_c$, the state of the system at Bob's mode collapses to
\begin{equation}
\begin{split}
    W_2(\chi_b &| q_a, p_c) = \mathcal{N} \int d p_a\, dq_c\, W_1(\chi_a, \chi_b, \chi_c) \\
    &= 2\mathcal{N} \int dx\, dy\, W_\text{in}\left(x, y \right) \\  &W_{a,b}\left(\sqrt{2}q_a - x, y - \sqrt{2}p_c, q_b, p_b \right) \text{ ,}
\end{split}
\end{equation}
where normalization must be added explicitly as a result of the non-unitary measurement operation, and we defined $\sqrt{2}x = q_a + q_c$ and $\sqrt{2}y = p_a + p_c$ for the last equality. If the entanglement resource is non-rotated two-mode squeezed vacuum, then Bob's output state approaches the exact input state in the infinite squeezing limit, aside from a phase space displacement, that is $W_2(\chi_b | q_a, p_c) \rightarrow W_\text{in}(q_b + \sqrt{2}q_a, p_b + \sqrt{2}p_c)$. To correct this, Bob displaces his mode, obtaining the final output state, given by the Wigner function
\begin{equation}
    W_\text{out}(q, p | q_a, p_c) = W_2(q - \sqrt{2}q_a, p - \sqrt{2}p_c | q_a, p_c) \text{ .}
\end{equation}
A scheme of the teleportation circuit can be seen in Fig.~1 of Ref.~\cite{braunstein_teleportation_1998}.

For the implementation of this protocol, we consider that the entangled two-mode squeezed state is sent to each party through a topological or trivial photonic lattice, as presented in section \ref{subsection:2mB}. Thus, at the output of lattice A, Alice extracts her share of the entangled state from the edge waveguide of the lattice, tracing out all other modes; and so does Bob with lattice B. 
Finally, we take the input state as a single-photon, $|\psi_\text{in}\rangle = c^\dag |0\rangle$. Note however, that the teleportation protocol is valid for any input, which might even be unknown to our protagonists.

\begin{figure*}[t]
    \centering
    \includegraphics[width=2\columnwidth]{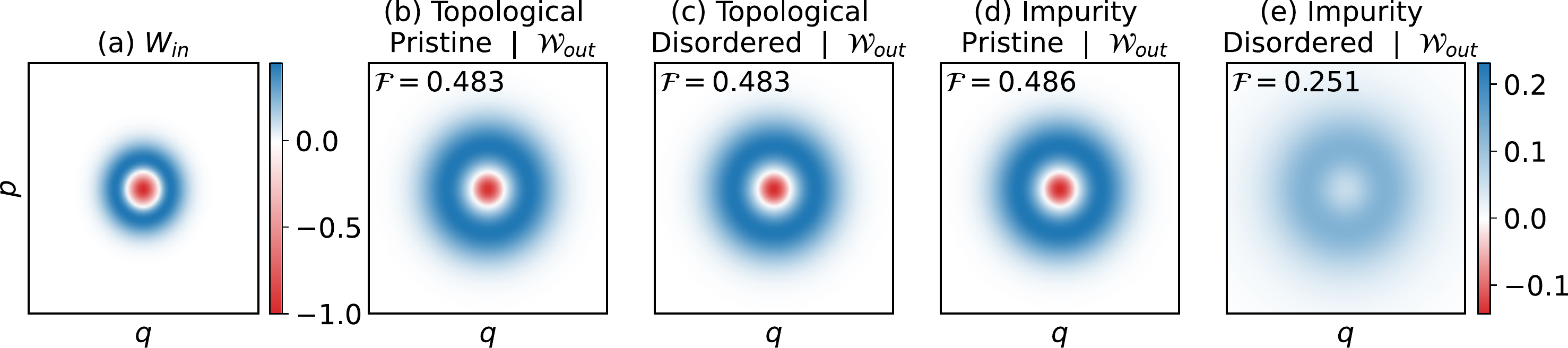}
    \caption{Panel (a) shows input Wigner function to be teleported, $W_\text{in}$. Panels (b) and (d) ((c) and (e)) show the averaged output Wigner function $\mathcal{W}_\text{out}$ for the pristine and hopping disordered $d=0.3v$ topological (topologically trivial) lattices. Wigner functions are expressed in units relative to the peak value of $W_\text{in}$. The fidelity of the output states is indicated on each panel, and the colorbar on the right applies for panels (b)--(e). Teleportation in the topological system proves to be robust against hopping disorder, preserving the fidelity of the output state, and preserving the quantum nature of the single-photon input state represented by negative values in the Wigner function.}
     \label{fig:teleportation}
\end{figure*}

We perform the teleportation for pristine and hopping disordered lattices ($d=0.3v$). The fidelity of a single teleportation event is calculated by $F = \pi \int dq\, dp\, W_\text{in}(q, p) W_\text{out}(q, p | q_a, p_c)$. Since the output state depends on the measured values of $q_a$ and $p_c$, and on the random disorder present in the lattice, we report the average fidelity over all measurements, and all random realizations, that is
\begin{equation}
    \mathcal{F} = \frac{1}{N} \sum_{n=1}^{N} \int dq_a\, dp_c \mathcal{P} F \text{ ,}
\end{equation}
where $N=300$ is the total number of random realizations. We may construct an averaged output Wigner function $\mathcal{W}_\text{out}$ defined by: 
\begin{equation}
\begin{split}
    \mathcal{W}_\text{out}(q, p) = \frac{1}{N} \sum_{n=1}^{N} &\int dq_a\, dp_c\, \mathcal{P}(q_a, p_c) \\ &W_\text{out}(q, p | q_a, p_c) \text{ ,}
\end{split}
\label{eq:wignerout_avg}
\end{equation}
so the average fidelity is $\mathcal{F} = \pi \int dq\, dp\, W_\text{in} \mathcal{W}_\text{out}$. Even though the output state never takes the exact form of $\mathcal{W}_\text{out}$, it allows to visualize the proximity of the output state to the input one in the general case.

When teleporting the single photon state employing the pristine lattices, we find practically identical results between the topological and trivial lattices, as shown in Fig.~\ref{fig:teleportation}-(b) and (d), characterized by an average fidelity of $0.483$ and $0.486$ respectively. Comparing with the input Wigner function (Fig.~\ref{fig:teleportation}-(a)), we observe that the fidelity losses are mainly explained by a reduction of the peak values, and a consequent widening of the function. Negativity of the output function serves as a landmark of the preservation of the input state's quantum nature, which can only be achieved by means of quantum teleportation schemes. Quantitatively, the reduction of the negative peak at the origin with respect to the input state is of $13.9\%$ and $14.3\%$ in the topological and trivial systems respectively.

The averaged output state of the topological system remains practically unchanged when introducing hopping disorder, as shown Fig.~\ref{fig:teleportation}-(c), maintaining the fidelity of the pristine lattice. The output Wigner function continues to take negative values, with a peak reduction of $14.1\%$, proving that teleportation in the topological system robustly preserves the quantum coherence of the input state. In contrast, teleportation in the topologically trivial system rapidly deteriorates as disorder is added. Even for low levels of disorder, the average fidelity falls to 0.242, roughly half of that of the the pristine system, and even though the output function presents a local minimum at the origin, it does not take negative values.

The drastic difference in the teleportation results between the topological and trivial systems is a consequence of topological protection of the phase of the squeezed quadrature. Transmission of the entangled two-mode state to the parties through a topological state ensures that Alice measures the maximally entangled variables, regardless of the magnitude of hopping disorder present in the lattice, resulting in robust preservation of the quantum coherence of the input state.

\section{Final remarks}
\label{section:conclusions}

We have studied the propagation of squeezing and entanglement along a topological SSH waveguide array, focusing on the effects of the lattices topology in this phenomena. We found that topological ordering of the lattice robustly preserves the phase of the squeezed quadrature when propagating squeezed light through a topologically protected localized state, for both single- and two-mode squeezed light. We also showed that the distribution of squeezing across the lattice inherits the spatial distribution of the localized state to which it is coupled, inheriting its topological protection as well. We discussed the interplay between entanglement, squeezing, and the phase of the squeezing; concluding that the topological phase of the lattice fixes the entangled variables. We finally provided a practical implementation of this system in a quantum teleportation protocol, where transmission of the entanglement resource to the parties through a topological lattice grants robustness and preservation of the quantum coherence of the teleported state.

We expect our findings on topological protection of squeezing to be of interest to any development in continuous variable quantum information, mainly as an alternative to protecting quantum coherence in photonic platforms. However, exploitation of topological phenomena in quantum optic systems, in particular continuous variable ones, is only recent, and its full possibilities are yet to be discovered. Our research is based on the SSH lattice, whose topological phase is broken by onsite disorder, which is always present to some extent in realistic devices; and being a one-dimensional lattice, this system is particularly sensitive to Anderson localization. Thus, a natural interest arises in studying the behavior of quantum light in two-dimensional lattices, with propagating edge states which enjoy better topological protection. 

\section{Acknowledgments}

We thank P. Solano for useful comments and suggestions. J.M.D. is supported by CONICYT Grant CONICYT-PFCHA/Mag\'{\i}sterNacional/2019-22200526. C.H-A. acknowledges support by CONICYT-PAI grant N°77180003, FONDECYT grant N°11190078, and ANID - Millenium Science Inititive Program - ICN17\_012. L.E.F.F.T. acknowledges the support of  FONDECYT grant N° 1211038, the Abdus Salam International Centre for Theoretical Physics and the Simons Foundation.


\begin{thebibliography}{57}%
\makeatletter
\providecommand \@ifxundefined [1]{%
 \@ifx{#1\undefined}
}%
\providecommand \@ifnum [1]{%
 \ifnum #1\expandafter \@firstoftwo
 \else \expandafter \@secondoftwo
 \fi
}%
\providecommand \@ifx [1]{%
 \ifx #1\expandafter \@firstoftwo
 \else \expandafter \@secondoftwo
 \fi
}%
\providecommand \natexlab [1]{#1}%
\providecommand \enquote  [1]{``#1''}%
\providecommand \bibnamefont  [1]{#1}%
\providecommand \bibfnamefont [1]{#1}%
\providecommand \citenamefont [1]{#1}%
\providecommand \href@noop [0]{\@secondoftwo}%
\providecommand \href [0]{\begingroup \@sanitize@url \@href}%
\providecommand \@href[1]{\@@startlink{#1}\@@href}%
\providecommand \@@href[1]{\endgroup#1\@@endlink}%
\providecommand \@sanitize@url [0]{\catcode `\\12\catcode `\$12\catcode
  `\&12\catcode `\#12\catcode `\^12\catcode `\_12\catcode `\%12\relax}%
\providecommand \@@startlink[1]{}%
\providecommand \@@endlink[0]{}%
\providecommand \url  [0]{\begingroup\@sanitize@url \@url }%
\providecommand \@url [1]{\endgroup\@href {#1}{\urlprefix }}%
\providecommand \urlprefix  [0]{URL }%
\providecommand \Eprint [0]{\href }%
\providecommand \doibase [0]{http://dx.doi.org/}%
\providecommand \selectlanguage [0]{\@gobble}%
\providecommand \bibinfo  [0]{\@secondoftwo}%
\providecommand \bibfield  [0]{\@secondoftwo}%
\providecommand \translation [1]{[#1]}%
\providecommand \BibitemOpen [0]{}%
\providecommand \bibitemStop [0]{}%
\providecommand \bibitemNoStop [0]{.\EOS\space}%
\providecommand \EOS [0]{\spacefactor3000\relax}%
\providecommand \BibitemShut  [1]{\csname bibitem#1\endcsname}%
\let\auto@bib@innerbib\@empty
\bibitem [{\citenamefont {von Klitzing}\ \emph {et~al.}(1980)\citenamefont {von
  Klitzing}, \citenamefont {Dorda},\ and\ \citenamefont
  {Pepper}}]{von_klitzing_new_1980}%
  \BibitemOpen
  \bibfield  {author} {\bibinfo {author} {\bibfnamefont {K.}~\bibnamefont {von
  Klitzing}}, \bibinfo {author} {\bibfnamefont {G.}~\bibnamefont {Dorda}}, \
  and\ \bibinfo {author} {\bibfnamefont {M.}~\bibnamefont {Pepper}},\
  }\bibfield  {title} {\enquote {\bibinfo {title} {New {Method} for
  {High}-{Accuracy} {Determination} of the {Fine}-{Structure} {Constant}
  {Based} on {Quantized} {Hall} {Resistance}},}\ }\href
  {http://link.aps.org/doi/10.1103/PhysRevLett.45.494} {\bibfield  {journal}
  {\bibinfo  {journal} {Phys. Rev. Lett.}\ }\textbf {\bibinfo {volume} {45}},\
  \bibinfo {pages} {494} (\bibinfo {year} {1980})}\BibitemShut {NoStop}%
\bibitem [{\citenamefont {Thouless}\ \emph {et~al.}(1982)\citenamefont
  {Thouless}, \citenamefont {Kohmoto}, \citenamefont {Nightingale},\ and\
  \citenamefont {den Nijs}}]{thouless_quantized_1982}%
  \BibitemOpen
  \bibfield  {author} {\bibinfo {author} {\bibfnamefont {D.~J.}\ \bibnamefont
  {Thouless}}, \bibinfo {author} {\bibfnamefont {M.}~\bibnamefont {Kohmoto}},
  \bibinfo {author} {\bibfnamefont {M.~P.}\ \bibnamefont {Nightingale}}, \ and\
  \bibinfo {author} {\bibfnamefont {M.}~\bibnamefont {den Nijs}},\ }\bibfield
  {title} {\enquote {\bibinfo {title} {Quantized {Hall} {Conductance} in a
  {Two}-{Dimensional} {Periodic} {Potential}},}\ }\href
  {http://link.aps.org/doi/10.1103/PhysRevLett.49.405} {\bibfield  {journal}
  {\bibinfo  {journal} {Phys. Rev. Lett.}\ }\textbf {\bibinfo {volume} {49}},\
  \bibinfo {pages} {405} (\bibinfo {year} {1982})}\BibitemShut {NoStop}%
\bibitem [{\citenamefont {Haldane}(1988)}]{haldane_model_1988}%
  \BibitemOpen
  \bibfield  {author} {\bibinfo {author} {\bibfnamefont {F.~D.~M.}\
  \bibnamefont {Haldane}},\ }\bibfield  {title} {\enquote {\bibinfo {title}
  {Model for a {Quantum} {Hall} {Effect} without {Landau} {Levels}:
  {Condensed}-{Matter} {Realization} of the "{Parity} {Anomaly}"},}\ }\href
  {http://link.aps.org/doi/10.1103/PhysRevLett.61.2015} {\bibfield  {journal}
  {\bibinfo  {journal} {Phys. Rev. Lett.}\ }\textbf {\bibinfo {volume} {61}},\
  \bibinfo {pages} {2015} (\bibinfo {year} {1988})}\BibitemShut {NoStop}%
\bibitem [{\citenamefont {Kane}\ and\ \citenamefont
  {Mele}(2005)}]{kane_quantum_2005}%
  \BibitemOpen
  \bibfield  {author} {\bibinfo {author} {\bibfnamefont {C.~L.}\ \bibnamefont
  {Kane}}\ and\ \bibinfo {author} {\bibfnamefont {E.~J.}\ \bibnamefont
  {Mele}},\ }\bibfield  {title} {\enquote {\bibinfo {title} {Quantum {Spin}
  {Hall} {Effect} in {Graphene}},}\ }\href
  {http://link.aps.org/doi/10.1103/PhysRevLett.95.226801} {\bibfield  {journal}
  {\bibinfo  {journal} {Phys. Rev. Lett.}\ }\textbf {\bibinfo {volume} {95}},\
  \bibinfo {pages} {226801} (\bibinfo {year} {2005})}\BibitemShut {NoStop}%
\bibitem [{\citenamefont {{C. L. Kane}}\ and\ \citenamefont {{M. Z.
  Hasan}}(2010)}]{c_l_kane_colloquium_2010}%
  \BibitemOpen
  \bibfield  {author} {\bibinfo {author} {\bibnamefont {{C. L. Kane}}}\ and\
  \bibinfo {author} {\bibnamefont {{M. Z. Hasan}}},\ }\bibfield  {title}
  {\enquote {\bibinfo {title} {Colloquium: {Topological} insulators},}\ }\href
  {\doibase 10.1103/RevModPhys.82.3045} {\bibfield  {journal} {\bibinfo
  {journal} {Rev. Mod. Phys}\ }\textbf {\bibinfo {volume} {82}},\ \bibinfo
  {pages} {3045} (\bibinfo {year} {2010})}\BibitemShut {NoStop}%
\bibitem [{\citenamefont {Ozawa}\ \emph {et~al.}(2019)\citenamefont {Ozawa},
  \citenamefont {Price}, \citenamefont {Amo}, \citenamefont {Goldman},
  \citenamefont {Hafezi}, \citenamefont {Lu}, \citenamefont {Rechtsman},
  \citenamefont {Schuster}, \citenamefont {Simon}, \citenamefont {Zilberberg},\
  and\ \citenamefont {Carusotto}}]{ozawa_topological_2019}%
  \BibitemOpen
  \bibfield  {author} {\bibinfo {author} {\bibfnamefont {T.}~\bibnamefont
  {Ozawa}}, \bibinfo {author} {\bibfnamefont {H.~M.}\ \bibnamefont {Price}},
  \bibinfo {author} {\bibfnamefont {A.}~\bibnamefont {Amo}}, \bibinfo {author}
  {\bibfnamefont {N.}~\bibnamefont {Goldman}}, \bibinfo {author} {\bibfnamefont
  {M.}~\bibnamefont {Hafezi}}, \bibinfo {author} {\bibfnamefont
  {L.}~\bibnamefont {Lu}}, \bibinfo {author} {\bibfnamefont {M.~C.}\
  \bibnamefont {Rechtsman}}, \bibinfo {author} {\bibfnamefont {D.}~\bibnamefont
  {Schuster}}, \bibinfo {author} {\bibfnamefont {J.}~\bibnamefont {Simon}},
  \bibinfo {author} {\bibfnamefont {O.}~\bibnamefont {Zilberberg}}, \ and\
  \bibinfo {author} {\bibfnamefont {I.}~\bibnamefont {Carusotto}},\ }\bibfield
  {title} {\enquote {\bibinfo {title} {Topological photonics},}\ }\href
  {\doibase 10.1103/RevModPhys.91.015006} {\bibfield  {journal} {\bibinfo
  {journal} {Reviews of Modern Physics}\ }\textbf {\bibinfo {volume} {91}},\
  \bibinfo {pages} {015006} (\bibinfo {year} {2019})}\BibitemShut {NoStop}%
\bibitem [{\citenamefont {Xie}\ \emph {et~al.}(2018)\citenamefont {Xie},
  \citenamefont {Wang}, \citenamefont {Zhu}, \citenamefont {Lu}, \citenamefont
  {Wang},\ and\ \citenamefont {Chen}}]{xie_photonics_2018}%
  \BibitemOpen
  \bibfield  {author} {\bibinfo {author} {\bibfnamefont {B.-Y.}\ \bibnamefont
  {Xie}}, \bibinfo {author} {\bibfnamefont {H.-F.}\ \bibnamefont {Wang}},
  \bibinfo {author} {\bibfnamefont {X.-Y.}\ \bibnamefont {Zhu}}, \bibinfo
  {author} {\bibfnamefont {M.-H.}\ \bibnamefont {Lu}}, \bibinfo {author}
  {\bibfnamefont {Z.~D.}\ \bibnamefont {Wang}}, \ and\ \bibinfo {author}
  {\bibfnamefont {Y.-F.}\ \bibnamefont {Chen}},\ }\bibfield  {title}
  {{
  \enquote {\bibinfo {title} {Photonics meets
  topology},}\ }}\href {\doibase 10.1364/OE.26.024531} {\bibfield  {journal}
  {\bibinfo  {journal} {Optics Express}\ }\textbf {\bibinfo {volume} {26}},\
  \bibinfo {pages} {24531} (\bibinfo {year} {2018})}\BibitemShut {NoStop}%
\bibitem [{\citenamefont {Lu}\ \emph {et~al.}(2014)\citenamefont {Lu},
  \citenamefont {Joannopoulos},\ and\ \citenamefont
  {Soljačić}}]{lu_topological_2014}%
  \BibitemOpen
  \bibfield  {author} {\bibinfo {author} {\bibfnamefont {L.}~\bibnamefont
  {Lu}}, \bibinfo {author} {\bibfnamefont {J.~D.}\ \bibnamefont
  {Joannopoulos}}, \ and\ \bibinfo {author} {\bibfnamefont {M.}~\bibnamefont
  {Soljačić}},\ }\bibfield  {title} {\enquote {\bibinfo {title} {Topological
  photonics},}\ }\href {\doibase 10.1038/nphoton.2014.248} {\bibfield
  {journal} {\bibinfo  {journal} {Nature Photonics}\ }\textbf {\bibinfo
  {volume} {8}},\ \bibinfo {pages} {821} (\bibinfo {year} {2014})}\BibitemShut
  {NoStop}%
\bibitem [{\citenamefont {Huber}(2016)}]{huber_topological_2016}%
  \BibitemOpen
  \bibfield  {author} {\bibinfo {author} {\bibfnamefont {S.~D.}\ \bibnamefont
  {Huber}},\ }\bibfield  {title} {{
  \enquote {\bibinfo
  {title} {Topological mechanics},}\ }}\href {\doibase 10.1038/nphys3801}
  {\bibfield  {journal} {\bibinfo  {journal} {Nature Physics}\ }\textbf
  {\bibinfo {volume} {12}},\ \bibinfo {pages} {621} (\bibinfo {year}
  {2016})}\BibitemShut {NoStop}%
\bibitem [{\citenamefont {Lee}\ \emph {et~al.}(2018)\citenamefont {Lee},
  \citenamefont {Imhof}, \citenamefont {Berger}, \citenamefont {Bayer},
  \citenamefont {Brehm}, \citenamefont {Molenkamp}, \citenamefont {Kiessling},\
  and\ \citenamefont {Thomale}}]{lee_topolectrical_2018}%
  \BibitemOpen
  \bibfield  {author} {\bibinfo {author} {\bibfnamefont {C.~H.}\ \bibnamefont
  {Lee}}, \bibinfo {author} {\bibfnamefont {S.}~\bibnamefont {Imhof}}, \bibinfo
  {author} {\bibfnamefont {C.}~\bibnamefont {Berger}}, \bibinfo {author}
  {\bibfnamefont {F.}~\bibnamefont {Bayer}}, \bibinfo {author} {\bibfnamefont
  {J.}~\bibnamefont {Brehm}}, \bibinfo {author} {\bibfnamefont {L.~W.}\
  \bibnamefont {Molenkamp}}, \bibinfo {author} {\bibfnamefont {T.}~\bibnamefont
  {Kiessling}}, \ and\ \bibinfo {author} {\bibfnamefont {R.}~\bibnamefont
  {Thomale}},\ }\bibfield  {title} {{\enquote {\bibinfo
  {title} {Topolectrical {Circuits}},}\ }}\href {https://doi.org/10.1038/s42005-018-0035-2} {\bibfield  {journal} {\bibinfo  {journal}
  {Communications Physics}\ }\textbf {\bibinfo {volume} {1}},\ \bibinfo {pages}
  {39} (\bibinfo {year} {2018})}\BibitemShut {NoStop}%
\bibitem [{\citenamefont {Peano}\ \emph {et~al.}(2015)\citenamefont {Peano},
  \citenamefont {Brendel}, \citenamefont {Schmidt},\ and\ \citenamefont
  {Marquardt}}]{peano_topological_2015}%
  \BibitemOpen
  \bibfield  {author} {\bibinfo {author} {\bibfnamefont {V.}~\bibnamefont
  {Peano}}, \bibinfo {author} {\bibfnamefont {C.}~\bibnamefont {Brendel}},
  \bibinfo {author} {\bibfnamefont {M.}~\bibnamefont {Schmidt}}, \ and\
  \bibinfo {author} {\bibfnamefont {F.}~\bibnamefont {Marquardt}},\ }\bibfield
  {title} {\enquote {\bibinfo {title} {Topological {Phases} of {Sound} and
  {Light}},}\ }\href {\doibase 10.1103/PhysRevX.5.031011} {\bibfield  {journal}
  {\bibinfo  {journal} {Physical Review X}\ }\textbf {\bibinfo {volume} {5}},\
  \bibinfo {pages} {031011} (\bibinfo {year} {2015})}\BibitemShut {NoStop}%
\bibitem [{\citenamefont {Khanikaev}\ \emph {et~al.}(2015)\citenamefont
  {Khanikaev}, \citenamefont {Fleury}, \citenamefont {Mousavi},\ and\
  \citenamefont {Alù}}]{khanikaev_topologically_2015}%
  \BibitemOpen
  \bibfield  {author} {\bibinfo {author} {\bibfnamefont {A.~B.}\ \bibnamefont
  {Khanikaev}}, \bibinfo {author} {\bibfnamefont {R.}~\bibnamefont {Fleury}},
  \bibinfo {author} {\bibfnamefont {S.~H.}\ \bibnamefont {Mousavi}}, \ and\
  \bibinfo {author} {\bibfnamefont {A.}~\bibnamefont {Alù}},\ }\bibfield
  {title} {\enquote {\bibinfo {title} {Topologically robust sound propagation
  in an angular-momentum-biased graphene-like resonator lattice},}\ }\href
  {\doibase 10.1038/ncomms9260} {\bibfield  {journal} {\bibinfo  {journal}
  {Nature Communications}\ }\textbf {\bibinfo {volume} {6}},\ \bibinfo {pages}
  {8260} (\bibinfo {year} {2015})}\BibitemShut {NoStop}%
\bibitem [{\citenamefont {Wang}\ \emph {et~al.}(2018)\citenamefont {Wang},
  \citenamefont {Zhang},\ and\ \citenamefont {Wang}}]{wang_topological_2018}%
  \BibitemOpen
  \bibfield  {author} {\bibinfo {author} {\bibfnamefont {X.}~\bibnamefont
  {Wang}}, \bibinfo {author} {\bibfnamefont {H.}~\bibnamefont {Zhang}}, \ and\
  \bibinfo {author} {\bibfnamefont {X.}~\bibnamefont {Wang}},\ }\bibfield
  {title} {\enquote {\bibinfo {title} {Topological {Magnonics}: {A} {Paradigm}
  for {Spin}-{Wave} {Manipulation} and {Device} {Design}},}\ }\href {https://doi.org/10.1103/PhysRevApplied.9.024029} {\bibfield  {journal} {\bibinfo  {journal}
  {Phys. Rev. Applied}\ }\textbf {\bibinfo {volume} {9}},\ \bibinfo {pages}
  {024029} (\bibinfo {year} {2018})},\BibitemShut {NoStop}%
\bibitem [{\citenamefont {Atala}\ \emph {et~al.}(2013)\citenamefont {Atala},
  \citenamefont {Aidelsburger}, \citenamefont {Barreiro}, \citenamefont
  {Abanin}, \citenamefont {Kitagawa}, \citenamefont {Demler},\ and\
  \citenamefont {Bloch}}]{atala_direct_2013}%
  \BibitemOpen
  \bibfield  {author} {\bibinfo {author} {\bibfnamefont {M.}~\bibnamefont
  {Atala}}, \bibinfo {author} {\bibfnamefont {M.}~\bibnamefont {Aidelsburger}},
  \bibinfo {author} {\bibfnamefont {J.~T.}\ \bibnamefont {Barreiro}}, \bibinfo
  {author} {\bibfnamefont {D.}~\bibnamefont {Abanin}}, \bibinfo {author}
  {\bibfnamefont {T.}~\bibnamefont {Kitagawa}}, \bibinfo {author}
  {\bibfnamefont {E.}~\bibnamefont {Demler}}, \ and\ \bibinfo {author}
  {\bibfnamefont {I.}~\bibnamefont {Bloch}},\ }\bibfield  {title}
  {{  \enquote {\bibinfo {title} {Direct measurement of the
  {Zak} phase in topological {Bloch} bands},}\ }}\href {https://doi.org/10.1038/nphys2790} {\bibfield  {journal} {\bibinfo  {journal} {Nature
  Physics}\ }\textbf {\bibinfo {volume} {9}},\ \bibinfo {pages} {795} (\bibinfo
  {year} {2013})}\BibitemShut {NoStop}%
\bibitem [{\citenamefont {Malkova}\ \emph {et~al.}(2009)\citenamefont
  {Malkova}, \citenamefont {Hromada}, \citenamefont {Wang}, \citenamefont
  {Bryant},\ and\ \citenamefont {Chen}}]{malkova_observation_2009}%
  \BibitemOpen
  \bibfield  {author} {\bibinfo {author} {\bibfnamefont {N.}~\bibnamefont
  {Malkova}}, \bibinfo {author} {\bibfnamefont {I.}~\bibnamefont {Hromada}},
  \bibinfo {author} {\bibfnamefont {X.}~\bibnamefont {Wang}}, \bibinfo {author}
  {\bibfnamefont {G.}~\bibnamefont {Bryant}}, \ and\ \bibinfo {author}
  {\bibfnamefont {Z.}~\bibnamefont {Chen}},\ }\bibfield  {title}
  {{
  \enquote {\bibinfo {title} {Observation of optical
  {Shockley}-like surface states in photonic superlattices},}\ }}\href
  {\doibase 10.1364/OL.34.001633} {\bibfield  {journal} {\bibinfo  {journal}
  {Optics Letters}\ }\textbf {\bibinfo {volume} {34}},\ \bibinfo {pages} {1633}
  (\bibinfo {year} {2009})}\BibitemShut {NoStop}%
\bibitem [{\citenamefont {König}\ \emph {et~al.}(2007)\citenamefont {König},
  \citenamefont {Wiedmann}, \citenamefont {Brüne}, \citenamefont {Roth},
  \citenamefont {Buhmann}, \citenamefont {Molenkamp}, \citenamefont {Qi},\ and\
  \citenamefont {Zhang}}]{konig_quantum_2007}%
  \BibitemOpen
  \bibfield  {author} {\bibinfo {author} {\bibfnamefont {M.}~\bibnamefont
  {König}}, \bibinfo {author} {\bibfnamefont {S.}~\bibnamefont {Wiedmann}},
  \bibinfo {author} {\bibfnamefont {C.}~\bibnamefont {Brüne}}, \bibinfo
  {author} {\bibfnamefont {A.}~\bibnamefont {Roth}}, \bibinfo {author}
  {\bibfnamefont {H.}~\bibnamefont {Buhmann}}, \bibinfo {author} {\bibfnamefont
  {L.~W.}\ \bibnamefont {Molenkamp}}, \bibinfo {author} {\bibfnamefont {X.-L.}\
  \bibnamefont {Qi}}, \ and\ \bibinfo {author} {\bibfnamefont {S.-C.}\
  \bibnamefont {Zhang}},\ }\bibfield  {title} {\enquote {\bibinfo {title}
  {Quantum {Spin} {Hall} {Insulator} {State} in {HgTe} {Quantum} {Wells}},}\
  }
  \href{\doibase 10.1126/science.1148047}
  {\bibfield
  {journal} {\bibinfo  {journal} {Science}\ }\textbf {\bibinfo {volume}
  {318}},\ \bibinfo {pages} {766} (\bibinfo {year} {2007})}\BibitemShut
  {NoStop}%
\bibitem [{\citenamefont {Hafezi}\ \emph {et~al.}(2011)\citenamefont {Hafezi},
  \citenamefont {Demler}, \citenamefont {Lukin},\ and\ \citenamefont
  {Taylor}}]{hafezi_robust_2011}%
  \BibitemOpen
  \bibfield  {author} {\bibinfo {author} {\bibfnamefont {M.}~\bibnamefont
  {Hafezi}}, \bibinfo {author} {\bibfnamefont {E.~A.}\ \bibnamefont {Demler}},
  \bibinfo {author} {\bibfnamefont {M.~D.}\ \bibnamefont {Lukin}}, \ and\
  \bibinfo {author} {\bibfnamefont {J.~M.}\ \bibnamefont {Taylor}},\ }\bibfield
   {title} {{
   \enquote {\bibinfo {title} {Robust optical
  delay lines with topological protection},}\ }}\href {https://doi.org/10.1038/nphys2063} {\bibfield  {journal} {\bibinfo  {journal} {Nature
  Physics}\ }\textbf {\bibinfo {volume} {7}},\ \bibinfo {pages} {907} (\bibinfo
  {year} {2011})}\BibitemShut {NoStop}%
\bibitem [{\citenamefont {Hsieh}\ \emph {et~al.}(2008)\citenamefont {Hsieh},
  \citenamefont {Qian}, \citenamefont {Wray}, \citenamefont {Xia},
  \citenamefont {Hor}, \citenamefont {Cava},\ and\ \citenamefont
  {Hasan}}]{hsieh_topological_2008}%
  \BibitemOpen
  \bibfield  {author} {\bibinfo {author} {\bibfnamefont {D.}~\bibnamefont
  {Hsieh}}, \bibinfo {author} {\bibfnamefont {D.}~\bibnamefont {Qian}},
  \bibinfo {author} {\bibfnamefont {L.}~\bibnamefont {Wray}}, \bibinfo {author}
  {\bibfnamefont {Y.}~\bibnamefont {Xia}}, \bibinfo {author} {\bibfnamefont
  {Y.~S.}\ \bibnamefont {Hor}}, \bibinfo {author} {\bibfnamefont {R.~J.}\
  \bibnamefont {Cava}}, \ and\ \bibinfo {author} {\bibfnamefont {M.~Z.}\
  \bibnamefont {Hasan}},\ }\bibfield  {title} {{
  \enquote
  {\bibinfo {title} {A topological {Dirac} insulator in a quantum spin {Hall}
  phase},}\ }}\href {\doibase 10.1038/nature06843} {\bibfield  {journal}
  {\bibinfo  {journal} {Nature}\ }\textbf {\bibinfo {volume} {452}},\ \bibinfo
  {pages} {970} (\bibinfo {year} {2008})}\BibitemShut {NoStop}%
\bibitem [{\citenamefont {Lu}\ \emph {et~al.}(2016)\citenamefont {Lu},
  \citenamefont {Fang}, \citenamefont {Fu}, \citenamefont {Johnson},
  \citenamefont {Joannopoulos},\ and\ \citenamefont
  {Soljačić}}]{lu_symmetry-protected_2016}%
  \BibitemOpen
  \bibfield  {author} {\bibinfo {author} {\bibfnamefont {L.}~\bibnamefont
  {Lu}}, \bibinfo {author} {\bibfnamefont {C.}~\bibnamefont {Fang}}, \bibinfo
  {author} {\bibfnamefont {L.}~\bibnamefont {Fu}}, \bibinfo {author}
  {\bibfnamefont {S.~G.}\ \bibnamefont {Johnson}}, \bibinfo {author}
  {\bibfnamefont {J.~D.}\ \bibnamefont {Joannopoulos}}, \ and\ \bibinfo
  {author} {\bibfnamefont {M.}~\bibnamefont {Soljačić}},\ }\bibfield  {title}
  {\enquote {\bibinfo {title} {Symmetry-protected topological photonic crystal
  in three dimensions},}\ }\href {\doibase 10.1038/nphys3611} {\bibfield
  {journal} {\bibinfo  {journal} {Nature Physics}\ }\textbf {\bibinfo {volume}
  {12}},\ \bibinfo {pages} {337} (\bibinfo {year} {2016})}\BibitemShut
  {NoStop}%
\bibitem [{\citenamefont {Hasan}\ \emph {et~al.}(2017)\citenamefont {Hasan},
  \citenamefont {Xu}, \citenamefont {Belopolski},\ and\ \citenamefont
  {Huang}}]{hasan_discovery_2017}%
  \BibitemOpen
  \bibfield  {author} {\bibinfo {author} {\bibfnamefont {M.~Z.}\ \bibnamefont
  {Hasan}}, \bibinfo {author} {\bibfnamefont {S.-Y.}\ \bibnamefont {Xu}},
  \bibinfo {author} {\bibfnamefont {I.}~\bibnamefont {Belopolski}}, \ and\
  \bibinfo {author} {\bibfnamefont {S.-M.}\ \bibnamefont {Huang}},\ }\bibfield
  {title} {\enquote {\bibinfo {title} {Discovery of {Weyl} {Fermion}
  {Semimetals} and {Topological} {Fermi} {Arc} {States}},}\ }\href {https://doi.org/10.1146/annurev-conmatphys-031016-025225} {\bibfield  {journal} {\bibinfo
  {journal} {Annual Review of Condensed Matter Physics}\ }\textbf {\bibinfo
  {volume} {8}},\ \bibinfo {pages} {289} (\bibinfo {year} {2017})}\BibitemShut
  {NoStop}%
\bibitem [{\citenamefont {Yang}\ \emph {et~al.}(2019)\citenamefont {Yang},
  \citenamefont {Gao}, \citenamefont {Xue}, \citenamefont {Zhang},
  \citenamefont {He}, \citenamefont {Yang}, \citenamefont {Singh},
  \citenamefont {Chong}, \citenamefont {Zhang},\ and\ \citenamefont
  {Chen}}]{yang_realization_2019}%
  \BibitemOpen
  \bibfield  {author} {\bibinfo {author} {\bibfnamefont {Y.}~\bibnamefont
  {Yang}}, \bibinfo {author} {\bibfnamefont {Z.}~\bibnamefont {Gao}}, \bibinfo
  {author} {\bibfnamefont {H.}~\bibnamefont {Xue}}, \bibinfo {author}
  {\bibfnamefont {L.}~\bibnamefont {Zhang}}, \bibinfo {author} {\bibfnamefont
  {M.}~\bibnamefont {He}}, \bibinfo {author} {\bibfnamefont {Z.}~\bibnamefont
  {Yang}}, \bibinfo {author} {\bibfnamefont {R.}~\bibnamefont {Singh}},
  \bibinfo {author} {\bibfnamefont {Y.}~\bibnamefont {Chong}}, \bibinfo
  {author} {\bibfnamefont {B.}~\bibnamefont {Zhang}}, \ and\ \bibinfo {author}
  {\bibfnamefont {H.}~\bibnamefont {Chen}},\ }\bibfield  {title} {\enquote
  {\bibinfo {title} {Realization of a three-dimensional photonic topological
  insulator},}\ }\href {\doibase 10.1038/s41586-018-0829-0} {\bibfield
  {journal} {\bibinfo  {journal} {Nature}\ }\textbf {\bibinfo {volume} {565}},\
  \bibinfo {pages} {622} (\bibinfo {year} {2019})}\BibitemShut {NoStop}%
\bibitem [{\citenamefont {Jotzu}\ \emph {et~al.}(2014)\citenamefont {Jotzu},
  \citenamefont {Messer}, \citenamefont {Desbuquois}, \citenamefont {Lebrat},
  \citenamefont {Uehlinger}, \citenamefont {Greif},\ and\ \citenamefont
  {Esslinger}}]{jotzu_experimental_2014}%
  \BibitemOpen
  \bibfield  {author} {\bibinfo {author} {\bibfnamefont {G.}~\bibnamefont
  {Jotzu}}, \bibinfo {author} {\bibfnamefont {M.}~\bibnamefont {Messer}},
  \bibinfo {author} {\bibfnamefont {R.}~\bibnamefont {Desbuquois}}, \bibinfo
  {author} {\bibfnamefont {M.}~\bibnamefont {Lebrat}}, \bibinfo {author}
  {\bibfnamefont {T.}~\bibnamefont {Uehlinger}}, \bibinfo {author}
  {\bibfnamefont {D.}~\bibnamefont {Greif}}, \ and\ \bibinfo {author}
  {\bibfnamefont {T.}~\bibnamefont {Esslinger}},\ }\bibfield  {title}
  {{ \enquote {\bibinfo {title} {Experimental realization of
  the topological {Haldane} model with ultracold fermions},}\ }}\href {https://doi.org/10.1038/nature13915} {\bibfield  {journal} {\bibinfo  {journal} {Nature}\
  }\textbf {\bibinfo {volume} {515}},\ \bibinfo {pages} {237} (\bibinfo {year}
  {2014})}\BibitemShut {NoStop}%
\bibitem [{\citenamefont {Haldane}\ and\ \citenamefont
  {Raghu}(2008)}]{haldane_possible_2008}%
  \BibitemOpen
  \bibfield  {author} {\bibinfo {author} {\bibfnamefont {F.~D.~M.}\
  \bibnamefont {Haldane}}\ and\ \bibinfo {author} {\bibfnamefont
  {S.}~\bibnamefont {Raghu}},\ }\bibfield  {title} {\enquote {\bibinfo {title}
  {Possible {Realization} of {Directional} {Optical} {Waveguides} in {Photonic}
  {Crystals} with {Broken} {Time}-{Reversal} {Symmetry}},}\ }\href {https://doi.org/10.1103/PhysRevLett.100.013904} {\bibfield  {journal} {\bibinfo  {journal}
  {Physical Review Letters}\ }\textbf {\bibinfo {volume} {100}},\ \bibinfo
  {pages} {013904} (\bibinfo {year} {2008})}\BibitemShut {NoStop}%
\bibitem [{\citenamefont {Raghu}\ and\ \citenamefont
  {Haldane}(2008)}]{raghu_analogs_2008}%
  \BibitemOpen
  \bibfield  {author} {\bibinfo {author} {\bibfnamefont {S.}~\bibnamefont
  {Raghu}}\ and\ \bibinfo {author} {\bibfnamefont {F.~D.~M.}\ \bibnamefont
  {Haldane}},\ }\bibfield  {title} {\enquote {\bibinfo {title} {Analogs of
  quantum-{Hall}-effect edge states in photonic crystals},}\ }\href {https://doi.org/10.1103/PhysRevA.78.033834} {\bibfield  {journal} {\bibinfo  {journal}
  {Physical Review A}\ }\textbf {\bibinfo {volume} {78}},\ \bibinfo {pages}
  {033834} (\bibinfo {year} {2008})}\BibitemShut {NoStop}%
\bibitem [{\citenamefont {Wang}\ \emph {et~al.}(2009)\citenamefont {Wang},
  \citenamefont {Chong}, \citenamefont {Joannopoulos},\ and\ \citenamefont
  {Soljačić}}]{wang_observation_2009}%
  \BibitemOpen
  \bibfield  {author} {\bibinfo {author} {\bibfnamefont {Z.}~\bibnamefont
  {Wang}}, \bibinfo {author} {\bibfnamefont {Y.}~\bibnamefont {Chong}},
  \bibinfo {author} {\bibfnamefont {J.~D.}\ \bibnamefont {Joannopoulos}}, \
  and\ \bibinfo {author} {\bibfnamefont {M.}~\bibnamefont {Soljačić}},\
  }\bibfield  {title} {{
  \enquote {\bibinfo {title}
  {Observation of unidirectional backscattering-immune topological
  electromagnetic states},}\ }}\href {\doibase 10.1038/nature08293} {\bibfield
  {journal} {\bibinfo  {journal} {Nature}\ }\textbf {\bibinfo {volume} {461}},\
  \bibinfo {pages} {772} (\bibinfo {year} {2009})}\BibitemShut {NoStop}%
\bibitem [{\citenamefont {Fu}\ and\ \citenamefont
  {Kane}(2007)}]{fu_topological_2007}%
  \BibitemOpen
  \bibfield  {author} {\bibinfo {author} {\bibfnamefont {L.}~\bibnamefont
  {Fu}}\ and\ \bibinfo {author} {\bibfnamefont {C.~L.}\ \bibnamefont {Kane}},\
  }\bibfield  {title} {\enquote {\bibinfo {title} {Topological insulators with
  inversion symmetry},}\ }\href {\doibase 10.1103/PhysRevB.76.045302}
  {\bibfield  {journal} {\bibinfo  {journal} {Physical Review B}\ }\textbf
  {\bibinfo {volume} {76}},\ \bibinfo {pages} {045302} (\bibinfo {year}
  {2007})}\BibitemShut {NoStop}%
\bibitem [{\citenamefont {Rudner}\ and\ \citenamefont
  {Lindner}(2020)}]{rudner_band_2020}%
  \BibitemOpen
  \bibfield  {author} {\bibinfo {author} {\bibfnamefont {M.~S.}\ \bibnamefont
  {Rudner}}\ and\ \bibinfo {author} {\bibfnamefont {N.~H.}\ \bibnamefont
  {Lindner}},\ }\bibfield  {title} {{
  \enquote {\bibinfo
  {title} {Band structure engineering and non-equilibrium dynamics in {Floquet}
  topological insulators},}\ }}\href {\doibase 10.1038/s42254-020-0170-z}
  {\bibfield  {journal} {\bibinfo  {journal} {Nature Reviews Physics}\ }\textbf
  {\bibinfo {volume} {2}},\ \bibinfo {pages} {229} (\bibinfo {year}
  {2020})}\BibitemShut {NoStop}%
\bibitem [{\citenamefont {Giustino}\ \emph {et~al.}(2020)\citenamefont
  {Giustino}, \citenamefont {Bibes}, \citenamefont {Lee}, \citenamefont
  {Trier}, \citenamefont {Valentí}, \citenamefont {Winter}, \citenamefont
  {Son}, \citenamefont {Taillefer}, \citenamefont {Heil}, \citenamefont
  {Figueroa}, \citenamefont {Plaçais}, \citenamefont {Wu}, \citenamefont
  {Yazyev}, \citenamefont {Bakkers}, \citenamefont {Nygård}, \citenamefont
  {Forn-Díaz}, \citenamefont {de~Franceschi}, \citenamefont {Foa~Torres},
  \citenamefont {McIver}, \citenamefont {Kumar}, \citenamefont {Low},
  \citenamefont {Galceran}, \citenamefont {Valenzuela}, \citenamefont
  {Costache}, \citenamefont {Manchon}, \citenamefont {Kim}, \citenamefont
  {Schleder}, \citenamefont {Fazzio},\ and\ \citenamefont
  {Roche}}]{giustino_2020_2020}%
  \BibitemOpen
  \bibfield  {author} {\bibinfo {author} {\bibfnamefont {F.}~\bibnamefont
  {Giustino}}, \bibinfo {author} {\bibfnamefont {M.}~\bibnamefont {Bibes}},
  \bibinfo {author} {\bibfnamefont {J.~H.}\ \bibnamefont {Lee}}, \bibinfo
  {author} {\bibfnamefont {F.}~\bibnamefont {Trier}}, \bibinfo {author}
  {\bibfnamefont {R.}~\bibnamefont {Valentí}}, \bibinfo {author}
  {\bibfnamefont {S.~M.}\ \bibnamefont {Winter}}, \bibinfo {author}
  {\bibfnamefont {Y.-W.}\ \bibnamefont {Son}}, \bibinfo {author} {\bibfnamefont
  {L.}~\bibnamefont {Taillefer}}, \bibinfo {author} {\bibfnamefont
  {C.}~\bibnamefont {Heil}}, \bibinfo {author} {\bibfnamefont {A.~I.}\
  \bibnamefont {Figueroa}}, \bibinfo {author} {\bibfnamefont {B.}~\bibnamefont
  {Plaçais}}, \bibinfo {author} {\bibfnamefont {Q.}~\bibnamefont {Wu}},
  \bibinfo {author} {\bibfnamefont {O.~V.}\ \bibnamefont {Yazyev}}, \bibinfo
  {author} {\bibfnamefont {E.~P. A.~M.}\ \bibnamefont {Bakkers}}, \bibinfo
  {author} {\bibfnamefont {J.}~\bibnamefont {Nygård}}, \bibinfo {author}
  {\bibfnamefont {P.}~\bibnamefont {Forn-Díaz}}, \bibinfo {author}
  {\bibfnamefont {S.}~\bibnamefont {de~Franceschi}}, \bibinfo {author}
  {\bibfnamefont {L.~E.~F.}\ \bibnamefont {Foa~Torres}}, \bibinfo {author}
  {\bibfnamefont {J.}~\bibnamefont {McIver}}, \bibinfo {author} {\bibfnamefont
  {A.}~\bibnamefont {Kumar}}, \bibinfo {author} {\bibfnamefont
  {T.}~\bibnamefont {Low}}, \bibinfo {author} {\bibfnamefont {R.}~\bibnamefont
  {Galceran}}, \bibinfo {author} {\bibfnamefont {S.~O.}\ \bibnamefont
  {Valenzuela}}, \bibinfo {author} {\bibfnamefont {M.~V.}\ \bibnamefont
  {Costache}}, \bibinfo {author} {\bibfnamefont {A.}~\bibnamefont {Manchon}},
  \bibinfo {author} {\bibfnamefont {E.-A.}\ \bibnamefont {Kim}}, \bibinfo
  {author} {\bibfnamefont {G.~R.}\ \bibnamefont {Schleder}}, \bibinfo {author}
  {\bibfnamefont {A.}~\bibnamefont {Fazzio}}, \ and\ \bibinfo {author}
  {\bibfnamefont {S.}~\bibnamefont {Roche}},\ }\bibfield  {title}
  {{ \enquote {\bibinfo {title} {The 2020 {Quantum}
  {Materials} {Roadmap}},}\ }}\href {\doibase 10.1088/2515-7639/abb74e}
  {\bibfield  {journal} {\bibinfo  {journal} {Journal of Physics: Materials}\ } \textbf {\bibinfo {volume} {3}},\ \bibinfo {pages} {014002}
  (\bibinfo {year} {2020})}\BibitemShut {NoStop}%
\bibitem [{\citenamefont {Rechtsman}\ \emph {et~al.}(2013)\citenamefont
  {Rechtsman}, \citenamefont {Zeuner}, \citenamefont {Plotnik}, \citenamefont
  {Lumer}, \citenamefont {Podolsky}, \citenamefont {Dreisow}, \citenamefont
  {Nolte}, \citenamefont {Segev},\ and\ \citenamefont
  {Szameit}}]{rechtsman_photonic_2013}%
  \BibitemOpen
  \bibfield  {author} {\bibinfo {author} {\bibfnamefont {M.~C.}\ \bibnamefont
  {Rechtsman}}, \bibinfo {author} {\bibfnamefont {J.~M.}\ \bibnamefont
  {Zeuner}}, \bibinfo {author} {\bibfnamefont {Y.}~\bibnamefont {Plotnik}},
  \bibinfo {author} {\bibfnamefont {Y.}~\bibnamefont {Lumer}}, \bibinfo
  {author} {\bibfnamefont {D.}~\bibnamefont {Podolsky}}, \bibinfo {author}
  {\bibfnamefont {F.}~\bibnamefont {Dreisow}}, \bibinfo {author} {\bibfnamefont
  {S.}~\bibnamefont {Nolte}}, \bibinfo {author} {\bibfnamefont
  {M.}~\bibnamefont {Segev}}, \ and\ \bibinfo {author} {\bibfnamefont
  {A.}~\bibnamefont {Szameit}},\ }\bibfield  {title} {\enquote {\bibinfo
  {title} {Photonic {Floquet} topological insulators},}\ }
  \href{\doibase 10.1038/nature12066}
  {\bibfield  {journal} {\bibinfo
  {journal} {Nature}\ }\textbf {\bibinfo {volume} {496}},\ \bibinfo {pages}
  {196} (\bibinfo {year} {2013})}\BibitemShut {NoStop}%
\bibitem [{\citenamefont {Foa~Torres}(2019)}]{foa_torres_perspective_2019}%
  \BibitemOpen
  \bibfield  {author} {\bibinfo {author} {\bibfnamefont {L.~E.~F.}\
  \bibnamefont {Foa~Torres}},\ }\bibfield  {title} {{
  \enquote {\bibinfo {title} {Perspective on topological states of
  non-{Hermitian} lattices},}\ }}\href {\doibase 10.1088/2515-7639/ab4092}
  {\bibfield  {journal} {\bibinfo  {journal} {Journal of Physics: Materials}\
  }\textbf {\bibinfo {volume} {3}},\ \bibinfo {pages} {014002} (\bibinfo {year}
  {2019})}\BibitemShut {NoStop}%
\bibitem [{\citenamefont {Weimann}\ \emph {et~al.}(2017)\citenamefont
  {Weimann}, \citenamefont {Kremer}, \citenamefont {Plotnik}, \citenamefont
  {Lumer}, \citenamefont {Nolte}, \citenamefont {Makris}, \citenamefont
  {Segev}, \citenamefont {Rechtsman},\ and\ \citenamefont
  {Szameit}}]{weimann_topologically_2017}%
  \BibitemOpen
  \bibfield  {author} {\bibinfo {author} {\bibfnamefont {S.}~\bibnamefont
  {Weimann}}, \bibinfo {author} {\bibfnamefont {M.}~\bibnamefont {Kremer}},
  \bibinfo {author} {\bibfnamefont {Y.}~\bibnamefont {Plotnik}}, \bibinfo
  {author} {\bibfnamefont {Y.}~\bibnamefont {Lumer}}, \bibinfo {author}
  {\bibfnamefont {S.}~\bibnamefont {Nolte}}, \bibinfo {author} {\bibfnamefont
  {K.~G.}\ \bibnamefont {Makris}}, \bibinfo {author} {\bibfnamefont
  {M.}~\bibnamefont {Segev}}, \bibinfo {author} {\bibfnamefont
  {M.}~\bibnamefont {Rechtsman}}, \ and\ \bibinfo {author} {\bibfnamefont
  {A.}~\bibnamefont {Szameit}},\ }\bibfield  {title} {\enquote {\bibinfo
  {title} {Topologically protected bound states in photonic
  parity–time-symmetric crystals},}\ }\href {\doibase 10.1038/nmat4811}
  {\bibfield  {journal} {\bibinfo  {journal} {Nature Materials}\ }\textbf
  {\bibinfo {volume} {16}},\ \bibinfo {pages} {433} (\bibinfo {year}
  {2017})}\BibitemShut {NoStop}%
\bibitem [{\citenamefont{Zhao}(2018)}]{zhao_topological_2018}%
  \BibitemOpen
  \bibfield  {author} {\bibinfo {author} {\bibfnamefont {H.}~\bibnamefont{Zhao}},\ \bibinfo {author} {\bibfnamefont {P.}~\bibnamefont{Miao}},\ \bibinfo {author} {\bibfnamefont {M.}~\bibnamefont{Teimourpour}},\ \bibinfo {author} {\bibfnamefont {S.}~\bibnamefont{Malzard}},\ \bibinfo {author} {\bibfnamefont {R.}~\bibnamefont{El-Ganainy}},\ \bibinfo {author} {\bibfnamefont {H.}~\bibnamefont{Schomerus}},\ \bibinfo {author} and {\bibfnamefont {L.}~\bibnamefont{Feng}},\ }
  \bibfield  {title} {\enquote {\bibinfo {title} {Topological hybrid silicon microlasers},}\ }\href
  {\doibase 10.1038/s41467-018-03434-2} {\bibfield  {journal} {\bibinfo
  {journal} {Nat. Commun.}\ }\textbf {\bibinfo {volume} {9}},\
  \bibinfo {pages} {981} (\bibinfo {year} {2018})}\BibitemShut {NoStop}%
\bibitem [{\citenamefont{St-Jean}(2017)}]{st-jean_lasing_2017}
  \BibitemOpen
  \bibfield  {author} {\bibinfo {author} {\bibfnamefont {P.}~\bibnamefont{St-Jean}},\ \bibinfo {author} {\bibfnamefont {V.}~\bibnamefont{Goblot}},\ \bibinfo {author} {\bibfnamefont {E.}~\bibnamefont{Galopin}},\ \bibinfo {author} {\bibfnamefont {A.}~\bibnamefont{Lema\^{i}tre}},\ \bibinfo {author} {\bibfnamefont {T.}~\bibnamefont{Ozawa}},\ \bibinfo {author} {\bibfnamefont {L.}~\bibnamefont{Le Gratiet}},\ \bibinfo {author} {\bibfnamefont {I.}~\bibnamefont{Sagnes}},\ \bibinfo {author} {\bibfnamefont {J.}~\bibnamefont{Bloch}},\ and \bibinfo {author} {\bibfnamefont {A.}~\bibnamefont{Amo}},\ }
  \bibfield  {title} {\enquote {\bibinfo {title} {Lasing in topological edge states of a one-dimensional lattice},}\ }\href
  {\doibase 10.1038/s41566-017-0006-2} {\bibfield  {journal} {\bibinfo
  {journal} {Nature Photon.}\ }\textbf {\bibinfo {volume} {11}},\
  \bibinfo {pages} {651} (\bibinfo {year} {2017})}\BibitemShut {NoStop}%
\bibitem [{\citenamefont{Peano}(2016)}]{peano_topological_2016-1}
  \BibitemOpen
  \bibfield  {author} {\bibinfo {author} {\bibfnamefont {V.}~\bibnamefont{Peano}},\ \bibinfo {author} {\bibfnamefont {M.}~\bibnamefont{Houde}},\ \bibinfo {author} {\bibfnamefont {F.}~\bibnamefont{Marquardt}},\ and \bibinfo {author} {\bibfnamefont {A.~A.}~\bibnamefont{Clerk}},\ }
  \bibfield  {title} {\enquote {\bibinfo {title} {Topological Quantum Fluctuations and Travelling Wave Amplifiers},}\ }\href
  {\doibase 10.1103/PhysRevX.6.041026} {\bibfield  {journal} {\bibinfo
  {journal} {Phys. Rev. X}\ }\textbf {\bibinfo {volume} {6}},\
  \bibinfo {pages} {041026} (\bibinfo {year} {2016})}\BibitemShut {NoStop}%
\bibitem [{\citenamefont{Wanjura}(2020)}]{wanjura_topological_2020}
  \BibitemOpen
  \bibfield  {author} {\bibinfo {author} {\bibfnamefont {C.}~\bibnamefont{Wanjura}},\ \bibinfo {author} {\bibfnamefont {M.}~\bibnamefont{Brunelli}},\ and \bibinfo {author} {\bibfnamefont {A.}~\bibnamefont{Nunnenkamp}},\ }
  \bibfield  {title} {\enquote {\bibinfo {title} {Topological framework for directional amplification in driven-dissipative cavity arrays},}\ }\href
  {\doibase 10.1038/s41467-020-16863-9} {\bibfield  {journal} {\bibinfo
  {journal} {Nat. Commun.}\ }\textbf {\bibinfo {volume} {11}},\
  \bibinfo {pages} {3149} (\bibinfo {year} {2020})}\BibitemShut {NoStop}%
\bibitem [{\citenamefont
  {Blanco-Redondo}(2020)}]{blanco-redondo_topological_2020}%
  \BibitemOpen
  \bibfield  {author} {\bibinfo {author} {\bibfnamefont {A.}~\bibnamefont
  {Blanco-Redondo}},\ }\bibfield  {title} {\enquote {\bibinfo {title}
  {Topological {Nanophotonics}: {Toward} {Robust} {Quantum} {Circuits}},}\
  }\href {\doibase 10.1109/JPROC.2019.2939987} {\bibfield  {journal} {\bibinfo
  {journal} {Proceedings of the IEEE}\ }\textbf {\bibinfo {volume} {108}},\
  \bibinfo {pages} {837} (\bibinfo {year} {2020})}\BibitemShut {NoStop}%
\bibitem [{\citenamefont {Kitagawa}\ \emph {et~al.}(2012)\citenamefont
  {Kitagawa}, \citenamefont {Broome}, \citenamefont {Fedrizzi}, \citenamefont
  {Rudner}, \citenamefont {Berg}, \citenamefont {Kassal}, \citenamefont
  {Aspuru-Guzik}, \citenamefont {Demler},\ and\ \citenamefont
  {White}}]{kitagawa_observation_2012}%
  \BibitemOpen
  \bibfield  {author} {\bibinfo {author} {\bibfnamefont {T.}~\bibnamefont
  {Kitagawa}}, \bibinfo {author} {\bibfnamefont {M.~A.}\ \bibnamefont
  {Broome}}, \bibinfo {author} {\bibfnamefont {A.}~\bibnamefont {Fedrizzi}},
  \bibinfo {author} {\bibfnamefont {M.~S.}\ \bibnamefont {Rudner}}, \bibinfo
  {author} {\bibfnamefont {E.}~\bibnamefont {Berg}}, \bibinfo {author}
  {\bibfnamefont {I.}~\bibnamefont {Kassal}}, \bibinfo {author} {\bibfnamefont
  {A.}~\bibnamefont {Aspuru-Guzik}}, \bibinfo {author} {\bibfnamefont
  {E.}~\bibnamefont {Demler}}, \ and\ \bibinfo {author} {\bibfnamefont {A.~G.}\
  \bibnamefont {White}},\ }\bibfield  {title} {\enquote {\bibinfo {title}
  {Observation of topologically protected bound states in photonic quantum
  walks},}\ }\href {\doibase 10.1038/ncomms1872} {\bibfield  {journal}
  {\bibinfo  {journal} {Nature Communications}\ }\textbf {\bibinfo {volume}
  {3}},\ \bibinfo {pages} {882} (\bibinfo {year} {2012})}\BibitemShut {NoStop}%
\bibitem [{\citenamefont {Chen}\ \emph {et~al.}(2018)\citenamefont {Chen},
  \citenamefont {Ding}, \citenamefont {Qin}, \citenamefont {He}, \citenamefont
  {Luo}, \citenamefont {Chen}, \citenamefont {Liu}, \citenamefont {Wang},
  \citenamefont {Zhang}, \citenamefont {Li}, \citenamefont {You}, \citenamefont
  {Wang}, \citenamefont {Wang}, \citenamefont {Sanders}, \citenamefont {Lu},\
  and\ \citenamefont {Pan}}]{chen_observation_2018}%
  \BibitemOpen
  \bibfield  {author} {\bibinfo {author} {\bibfnamefont {C.}~\bibnamefont
  {Chen}}, \bibinfo {author} {\bibfnamefont {X.}~\bibnamefont {Ding}}, \bibinfo
  {author} {\bibfnamefont {J.}~\bibnamefont {Qin}}, \bibinfo {author}
  {\bibfnamefont {Y.}~\bibnamefont {He}}, \bibinfo {author} {\bibfnamefont
  {Y.-H.}\ \bibnamefont {Luo}}, \bibinfo {author} {\bibfnamefont {M.-C.}\
  \bibnamefont {Chen}}, \bibinfo {author} {\bibfnamefont {C.}~\bibnamefont
  {Liu}}, \bibinfo {author} {\bibfnamefont {X.-L.}\ \bibnamefont {Wang}},
  \bibinfo {author} {\bibfnamefont {W.-J.}\ \bibnamefont {Zhang}}, \bibinfo
  {author} {\bibfnamefont {H.}~\bibnamefont {Li}}, \bibinfo {author}
  {\bibfnamefont {L.-X.}\ \bibnamefont {You}}, \bibinfo {author} {\bibfnamefont
  {Z.}~\bibnamefont {Wang}}, \bibinfo {author} {\bibfnamefont {D.-W.}\
  \bibnamefont {Wang}}, \bibinfo {author} {\bibfnamefont {B.~C.}\ \bibnamefont
  {Sanders}}, \bibinfo {author} {\bibfnamefont {C.-Y.}\ \bibnamefont {Lu}}, \
  and\ \bibinfo {author} {\bibfnamefont {J.-W.}\ \bibnamefont {Pan}},\
  }\bibfield  {title} {\enquote {\bibinfo {title} {Observation of
  {Topologically} {Protected} {Edge} {States} in a {Photonic}
  {Two}-{Dimensional} {Quantum} {Walk}},}\ }\href {https://doi.org/10.1103/PhysRevLett.121.100502} {\bibfield  {journal} {\bibinfo  {journal}
  {Phys. Rev. Lett.}\ }\textbf {\bibinfo {volume} {121}},\ \bibinfo {pages}
  {100502} (\bibinfo {year} {2018})}.
\bibitem [{\citenamefont {Barik}\ \emph {et~al.}(2018)\citenamefont {Barik},
  \citenamefont {Karasahin}, \citenamefont {Flower}, \citenamefont {Cai},
  \citenamefont {Miyake}, \citenamefont {DeGottardi}, \citenamefont {Hafezi},\
  and\ \citenamefont {Waks}}]{barik_topological_2018}%
  \BibitemOpen
  \bibfield  {author} {\bibinfo {author} {\bibfnamefont {S.}~\bibnamefont
  {Barik}}, \bibinfo {author} {\bibfnamefont {A.}~\bibnamefont {Karasahin}},
  \bibinfo {author} {\bibfnamefont {C.}~\bibnamefont {Flower}}, \bibinfo
  {author} {\bibfnamefont {T.}~\bibnamefont {Cai}}, \bibinfo {author}
  {\bibfnamefont {H.}~\bibnamefont {Miyake}}, \bibinfo {author} {\bibfnamefont
  {W.}~\bibnamefont {DeGottardi}}, \bibinfo {author} {\bibfnamefont
  {M.}~\bibnamefont {Hafezi}}, \ and\ \bibinfo {author} {\bibfnamefont
  {E.}~\bibnamefont {Waks}},\ }\bibfield  {title} {\enquote {\bibinfo {title}
  {A topological quantum optics interface},}\ }\href {https://doi.org/10.1126/science.aaq0327} {\bibfield  {journal} {\bibinfo  {journal}
  {Science}\ }\textbf {\bibinfo {volume} {359}},\ \bibinfo {pages} {666}
  (\bibinfo {year} {2018})}\BibitemShut {NoStop}%
\bibitem [{\citenamefont {Tambasco}\ \emph {et~al.}(2018)\citenamefont
  {Tambasco}, \citenamefont {Corrielli}, \citenamefont {Chapman}, \citenamefont
  {Crespi}, \citenamefont {Zilberberg}, \citenamefont {Osellame},\ and\
  \citenamefont {Peruzzo}}]{tambasco_quantum_2018}%
  \BibitemOpen
  \bibfield  {author} {\bibinfo {author} {\bibfnamefont {J.-L.}\ \bibnamefont
  {Tambasco}}, \bibinfo {author} {\bibfnamefont {G.}~\bibnamefont {Corrielli}},
  \bibinfo {author} {\bibfnamefont {R.~J.}\ \bibnamefont {Chapman}}, \bibinfo
  {author} {\bibfnamefont {A.}~\bibnamefont {Crespi}}, \bibinfo {author}
  {\bibfnamefont {O.}~\bibnamefont {Zilberberg}}, \bibinfo {author}
  {\bibfnamefont {R.}~\bibnamefont {Osellame}}, \ and\ \bibinfo {author}
  {\bibfnamefont {A.}~\bibnamefont {Peruzzo}},\ }\bibfield  {title}
  {{
  \enquote {\bibinfo {title} {Quantum interference of
  topological states of light},}\ }}\href {\doibase 10.1126/sciadv.aat3187}
  {\bibfield  {journal} {\bibinfo  {journal} {Science Advances}\ }\textbf
  {\bibinfo {volume} {4}},\ \bibinfo {pages} {eaat3187} (\bibinfo {year}
  {2018})}\BibitemShut {NoStop}%
\bibitem [{\citenamefont {Mittal}\ \emph {et~al.}(2018)\citenamefont {Mittal},
  \citenamefont {Goldschmidt},\ and\ \citenamefont
  {Hafezi}}]{mittal_topological_2018}%
  \BibitemOpen
  \bibfield  {author} {\bibinfo {author} {\bibfnamefont {S.}~\bibnamefont
  {Mittal}}, \bibinfo {author} {\bibfnamefont {E.~A.}\ \bibnamefont
  {Goldschmidt}}, \ and\ \bibinfo {author} {\bibfnamefont {M.}~\bibnamefont
  {Hafezi}},\ }\bibfield  {title} {{
  \enquote {\bibinfo
  {title} {A topological source of quantum light},}\ }}\href {https://doi.org/10.1038/s41586-018-0478-3} {\bibfield  {journal} {\bibinfo  {journal}
  {Nature}\ }\textbf {\bibinfo {volume} {561}},\ \bibinfo {pages} {502}
  (\bibinfo {year} {2018})}\BibitemShut {NoStop}%
\bibitem [{\citenamefont {Blanco-Redondo}\ \emph {et~al.}(2018)\citenamefont
  {Blanco-Redondo}, \citenamefont {Bell}, \citenamefont {Oren}, \citenamefont
  {Eggleton},\ and\ \citenamefont {Segev}}]{blanco-redondo_topological_2018}%
  \BibitemOpen
  \bibfield  {author} {\bibinfo {author} {\bibfnamefont {A.}~\bibnamefont
  {Blanco-Redondo}}, \bibinfo {author} {\bibfnamefont {B.}~\bibnamefont
  {Bell}}, \bibinfo {author} {\bibfnamefont {D.}~\bibnamefont {Oren}}, \bibinfo
  {author} {\bibfnamefont {B.~J.}\ \bibnamefont {Eggleton}}, \ and\ \bibinfo
  {author} {\bibfnamefont {M.}~\bibnamefont {Segev}},\ }\bibfield  {title}
  {{
  \enquote {\bibinfo {title} {Topological protection of
  biphoton states},}\ }}\href {\doibase 10.1126/science.aau4296} {\bibfield
  {journal} {\bibinfo  {journal} {Science}\ }\textbf {\bibinfo {volume}
  {362}},\ \bibinfo {pages} {568} (\bibinfo {year} {2018})}\BibitemShut
  {NoStop}%
\bibitem [{\citenamefont {Rechtsman}\ \emph {et~al.}(2016)\citenamefont
  {Rechtsman}, \citenamefont {Lumer}, \citenamefont {Plotnik}, \citenamefont
  {Perez-Leija}, \citenamefont {Szameit},\ and\ \citenamefont
  {Segev}}]{rechtsman_topological_2016}%
  \BibitemOpen
  \bibfield  {author} {\bibinfo {author} {\bibfnamefont {M.~C.}\ \bibnamefont
  {Rechtsman}}, \bibinfo {author} {\bibfnamefont {Y.}~\bibnamefont {Lumer}},
  \bibinfo {author} {\bibfnamefont {Y.}~\bibnamefont {Plotnik}}, \bibinfo
  {author} {\bibfnamefont {A.}~\bibnamefont {Perez-Leija}}, \bibinfo {author}
  {\bibfnamefont {A.}~\bibnamefont {Szameit}}, \ and\ \bibinfo {author}
  {\bibfnamefont {M.}~\bibnamefont {Segev}},\ }\bibfield  {title}
  {{
  \enquote {\bibinfo {title} {Topological protection of
  photonic path entanglement},}\ }}\href {\doibase 10.1364/OPTICA.3.000925}
  {\bibfield  {journal} {\bibinfo  {journal} {Optica}\ }\textbf {\bibinfo
  {volume} {3}},\ \bibinfo {pages} {925} (\bibinfo {year} {2016})}\BibitemShut
  {NoStop}%
\bibitem [{\citenamefont {Mittal}\ \emph {et~al.}(2016)\citenamefont {Mittal},
  \citenamefont {Orre},\ and\ \citenamefont
  {Hafezi}}]{mittal_topologically_2016}%
  \BibitemOpen
  \bibfield  {author} {\bibinfo {author} {\bibfnamefont {S.}~\bibnamefont
  {Mittal}}, \bibinfo {author} {\bibfnamefont {V.~V.}\ \bibnamefont {Orre}}, \
  and\ \bibinfo {author} {\bibfnamefont {M.}~\bibnamefont {Hafezi}},\
  }\bibfield  {title} {{
  \enquote {\bibinfo {title}
  {Topologically robust transport of entangled photons in a {2D} photonic
  system},}\ }}\href {\doibase 10.1364/OE.24.015631} {\bibfield  {journal}
  {\bibinfo  {journal} {Optics Express}\ }\textbf {\bibinfo {volume} {24}},\
  \bibinfo {pages} {15631} (\bibinfo {year} {2016})}\BibitemShut {NoStop}%
\bibitem [{\citenamefont {Wang}\ \emph {et~al.}(2019)\citenamefont {Wang},
  \citenamefont {Doyle}, \citenamefont {Bell}, \citenamefont {Collins},
  \citenamefont {Magi}, \citenamefont {Eggleton}, \citenamefont {Segev},\ and\
  \citenamefont {Blanco-Redondo}}]{wang_topologically_2019}%
  \BibitemOpen
  \bibfield  {author} {\bibinfo {author} {\bibfnamefont {M.}~\bibnamefont
  {Wang}}, \bibinfo {author} {\bibfnamefont {C.}~\bibnamefont {Doyle}},
  \bibinfo {author} {\bibfnamefont {B.}~\bibnamefont {Bell}}, \bibinfo {author}
  {\bibfnamefont {M.~J.}\ \bibnamefont {Collins}}, \bibinfo {author}
  {\bibfnamefont {E.}~\bibnamefont {Magi}}, \bibinfo {author} {\bibfnamefont
  {B.~J.}\ \bibnamefont {Eggleton}}, \bibinfo {author} {\bibfnamefont
  {M.}~\bibnamefont {Segev}}, \ and\ \bibinfo {author} {\bibfnamefont
  {A.}~\bibnamefont {Blanco-Redondo}},\ }\bibfield  {title} {{
  \enquote {\bibinfo {title} {Topologically protected entangled photonic
  states},}\ }}\href {\doibase 10.1515/nanoph-2019-0058} {\bibfield  {journal}
  {\bibinfo  {journal} {Nanophotonics}\ }\textbf {\bibinfo {volume} {8}},\
  \bibinfo {pages} {1327} (\bibinfo {year} {2019})}\BibitemShut {NoStop}%
\bibitem [{\citenamefont {Peano}(2016)}]{peano_topological_2016}%
  \BibitemOpen
  \bibfield  {author} {\bibinfo {author} {\bibfnamefont {V.}~\bibnamefont{Peano}},\ \bibinfo {author} {\bibfnamefont {M.}~\bibnamefont{Houde}},\ \bibinfo {author} {\bibfnamefont {C.}~\bibnamefont{Brendel}},\ \bibinfo {author} {\bibfnamefont {F.}~\bibnamefont{Marquardt}},\ and \bibinfo {author} {\bibfnamefont {A.~A.}~\bibnamefont{Clerk}},\ }
  \bibfield  {title} {\enquote {\bibinfo {title} {Topological phase transitions and chiral inelastic transport induced by the squeezing of light},}\ }\href
  {\doibase 10.1038/ncomms10779} {\bibfield  {journal} {\bibinfo
  {journal} {Nature Communications}\ }\textbf {\bibinfo {volume} {7}},\
  \bibinfo {pages} {10779} (\bibinfo {year} {2016})}\BibitemShut {NoStop}%
\bibitem [{\citenamefont {Braunstein}\ and\ \citenamefont {van
  Loock}(2005)}]{braunstein_quantum_2005}%
  \BibitemOpen
  \bibfield  {author} {\bibinfo {author} {\bibfnamefont {S.~L.}\ \bibnamefont
  {Braunstein}}\ and\ \bibinfo {author} {\bibfnamefont {P.}~\bibnamefont {van
  Loock}},\ }\bibfield  {title} {\enquote {\bibinfo {title} {Quantum
  information with continuous variables},}\ }\href {https://doi.org/10.1103/RevModPhys.77.513} {\bibfield  {journal} {\bibinfo  {journal}
  {Reviews of Modern Physics}\ }\textbf {\bibinfo {volume} {77}},\ \bibinfo
  {pages} {513} (\bibinfo {year} {2005})}\BibitemShut {NoStop}%
\bibitem [{\citenamefont {Gottesman}\ \emph {et~al.}(2001)\citenamefont
  {Gottesman}, \citenamefont {Kitaev},\ and\ \citenamefont
  {Preskill}}]{gottesman_encoding_2001}%
  \BibitemOpen
  \bibfield  {author} {\bibinfo {author} {\bibfnamefont {D.}~\bibnamefont
  {Gottesman}}, \bibinfo {author} {\bibfnamefont {A.}~\bibnamefont {Kitaev}}, \
  and\ \bibinfo {author} {\bibfnamefont {J.}~\bibnamefont {Preskill}},\
  }\bibfield  {title} {\enquote {\bibinfo {title} {Encoding a qubit in an
  oscillator},}\ }\href {\doibase 10.1103/PhysRevA.64.012310} {\bibfield
  {journal} {\bibinfo  {journal} {Physical Review A}\ }\textbf {\bibinfo
  {volume} {64}},\ \bibinfo {pages} {012310} (\bibinfo {year}
  {2001})}\BibitemShut {NoStop}%
\bibitem [{\citenamefont {Menicucci}\ \emph {et~al.}(2006)\citenamefont
  {Menicucci}, \citenamefont {van Loock}, \citenamefont {Gu}, \citenamefont
  {Weedbrook}, \citenamefont {Ralph},\ and\ \citenamefont
  {Nielsen}}]{menicucci_universal_2006}%
  \BibitemOpen
  \bibfield  {author} {\bibinfo {author} {\bibfnamefont {N.~C.}\ \bibnamefont
  {Menicucci}}, \bibinfo {author} {\bibfnamefont {P.}~\bibnamefont {van
  Loock}}, \bibinfo {author} {\bibfnamefont {M.}~\bibnamefont {Gu}}, \bibinfo
  {author} {\bibfnamefont {C.}~\bibnamefont {Weedbrook}}, \bibinfo {author}
  {\bibfnamefont {T.~C.}\ \bibnamefont {Ralph}}, \ and\ \bibinfo {author}
  {\bibfnamefont {M.~A.}\ \bibnamefont {Nielsen}},\ }\bibfield  {title}
  {\enquote {\bibinfo {title} {Universal {Quantum} {Computation} with
  {Continuous}-{Variable} {Cluster} {States}},}\ }\href {https://doi.org/10.1103/PhysRevLett.97.110501} {\bibfield  {journal} {\bibinfo  {journal}
  {Phys. Rev. Lett.}\ }\textbf {\bibinfo {volume} {97}},\ \bibinfo {pages}
  {110501} (\bibinfo {year} {2006})}.
\bibitem [{\citenamefont {Caves}(1981)}]{caves_quantum-mechanical_1981}%
  \BibitemOpen
  \bibfield  {author} {\bibinfo {author} {\bibfnamefont {C.~M.}\ \bibnamefont
  {Caves}},\ }\bibfield  {title} {\enquote {\bibinfo {title}
  {Quantum-mechanical noise in an interferometer},}\ }\href {https://doi.org/10.1103/PhysRevD.23.1693} {\bibfield  {journal} {\bibinfo  {journal} {Phys.
  Rev. D}\ }\textbf {\bibinfo {volume} {23}},\ \bibinfo {pages} {1693}
  (\bibinfo {year} {1981})}.
\bibitem [{\citenamefont {Aasi}\ and\ \citenamefont {the LIGO
  Scientific~Collaboration}(2013)}]{aasi_enhanced_2013}%
  \BibitemOpen
  \bibfield  {author} {\bibinfo {author} {\bibfnamefont {J.}~\bibnamefont
  {Aasi}}\ and\ \bibinfo {author} {\bibnamefont {the LIGO
  Scientific~Collaboration}},\ }\bibfield  {title} {{
  \enquote {\bibinfo {title} {Enhanced sensitivity of the {LIGO}
  gravitational wave detector by using squeezed states of light},}\ }}\href
  {\doibase 10.1038/nphoton.2013.177} {\bibfield  {journal} {\bibinfo
  {journal} {Nature Photonics}\ }\textbf {\bibinfo {volume} {7}},\ \bibinfo
  {pages} {613} (\bibinfo {year} {2013})}\BibitemShut {NoStop}%
\bibitem [{\citenamefont {Larsen}\ \emph {et~al.}(2019)\citenamefont {Larsen},
  \citenamefont {Guo}, \citenamefont {Breum}, \citenamefont
  {Neergaard-Nielsen},\ and\ \citenamefont
  {Andersen}}]{larsen_deterministic_2019}%
  \BibitemOpen
  \bibfield  {author} {\bibinfo {author} {\bibfnamefont {M.~V.}\ \bibnamefont
  {Larsen}}, \bibinfo {author} {\bibfnamefont {X.}~\bibnamefont {Guo}},
  \bibinfo {author} {\bibfnamefont {C.~R.}\ \bibnamefont {Breum}}, \bibinfo
  {author} {\bibfnamefont {J.~S.}\ \bibnamefont {Neergaard-Nielsen}}, \ and\
  \bibinfo {author} {\bibfnamefont {U.~L.}\ \bibnamefont {Andersen}},\
  }\bibfield  {title} {\enquote {\bibinfo {title} {Deterministic generation of
  a two-dimensional cluster state},}\ }\href {\doibase 10.1126/science.aay4354}
  {\bibfield  {journal} {\bibinfo  {journal} {Science}\ }\textbf {\bibinfo
  {volume} {366}},\ \bibinfo {pages} {369} (\bibinfo {year} {2019})}.
\bibitem [{\citenamefont {Braunstein}\ and\ \citenamefont
  {Kimble}(1998)}]{braunstein_teleportation_1998}%
  \BibitemOpen
  \bibfield  {author} {\bibinfo {author} {\bibfnamefont {S.~L.}\ \bibnamefont
  {Braunstein}}\ and\ \bibinfo {author} {\bibfnamefont {H.~J.}\ \bibnamefont
  {Kimble}},\ }\bibfield  {title} {\enquote {\bibinfo {title} {Teleportation of
  {Continuous} {Quantum} {Variables}},}\ }\href {https://doi.org/10.1103/PhysRevLett.80.869} {\bibfield  {journal} {\bibinfo  {journal} {Phys.
  Rev. Lett.}\ }\textbf {\bibinfo {volume} {80}},\ \bibinfo {pages} {869}
  (\bibinfo {year} {1998})}.
\bibitem [{\citenamefont {Yukawa}\ \emph {et~al.}(2008)\citenamefont {Yukawa},
  \citenamefont {Benichi},\ and\ \citenamefont
  {Furusawa}}]{yukawa_high-fidelity_2008}%
  \BibitemOpen
  \bibfield  {author} {\bibinfo {author} {\bibfnamefont {M.}~\bibnamefont
  {Yukawa}}, \bibinfo {author} {\bibfnamefont {H.}~\bibnamefont {Benichi}}, \
  and\ \bibinfo {author} {\bibfnamefont {A.}~\bibnamefont {Furusawa}},\
  }\bibfield  {title} {\enquote {\bibinfo {title} {High-fidelity
  continuous-variable quantum teleportation toward multistep quantum
  operations},}\ }\href {\doibase 10.1103/PhysRevA.77.022314} {\bibfield
  {journal} {\bibinfo  {journal} {Phys. Rev. A}\ }\textbf {\bibinfo {volume}
  {77}},\ \bibinfo {pages} {022314} (\bibinfo {year} {2008})}.
\bibitem [{\citenamefont {Bourassa}\ \emph {et~al.}(2021)\citenamefont
  {Bourassa}, \citenamefont {Alexander}, \citenamefont {Vasmer}, \citenamefont
  {Patil}, \citenamefont {Tzitrin}, \citenamefont {Matsuura}, \citenamefont
  {Su}, \citenamefont {Baragiola}, \citenamefont {Guha}, \citenamefont
  {Dauphinais},\ and\ \citenamefont {al}}]{bourassa_blueprint_2021}%
  \BibitemOpen
  \bibfield  {author} {\bibinfo {author} {\bibfnamefont {J.~E.}\ \bibnamefont
  {Bourassa}}, \bibinfo {author} {\bibfnamefont {R.~N.}\ \bibnamefont
  {Alexander}}, \bibinfo {author} {\bibfnamefont {M.}~\bibnamefont {Vasmer}},
  \bibinfo {author} {\bibfnamefont {A.}~\bibnamefont {Patil}}, \bibinfo
  {author} {\bibfnamefont {I.}~\bibnamefont {Tzitrin}}, \bibinfo {author}
  {\bibfnamefont {T.}~\bibnamefont {Matsuura}}, \bibinfo {author}
  {\bibfnamefont {D.}~\bibnamefont {Su}}, \bibinfo {author} {\bibfnamefont
  {B.~Q.}\ \bibnamefont {Baragiola}}, \bibinfo {author} {\bibfnamefont
  {S.}~\bibnamefont {Guha}}, \bibinfo {author} {\bibfnamefont {G.}~\bibnamefont
  {Dauphinais}}, \ and\ \bibinfo {author} {\bibfnamefont {e.}~\bibnamefont
  {al}},\ }\bibfield  {title} {\enquote {\bibinfo {title} {Blueprint for a
  {Scalable} {Photonic} {Fault}-{Tolerant} {Quantum} {Computer}},}\ }\href
  {\doibase 10.22331/q-2021-02-04-392} {\bibfield  {journal} {\bibinfo
  {journal} {Quantum}\ }\textbf {\bibinfo {volume} {5}},\ \bibinfo {pages}
  {392} (\bibinfo {year} {2021})}.
\bibitem [{\citenamefont {Larsen}\ \emph {et~al.}(2021)\citenamefont {Larsen},
  \citenamefont {Chamberland}, \citenamefont {Noh}, \citenamefont
  {Neergaard-Nielsen},\ and\ \citenamefont
  {Andersen}}]{larsen_fault-tolerant_2021}%
  \BibitemOpen
  \bibfield  {author} {\bibinfo {author} {\bibfnamefont {M.~V.}\ \bibnamefont
  {Larsen}}, \bibinfo {author} {\bibfnamefont {C.}~\bibnamefont {Chamberland}},
  \bibinfo {author} {\bibfnamefont {K.}~\bibnamefont {Noh}}, \bibinfo {author}
  {\bibfnamefont {J.~S.}\ \bibnamefont {Neergaard-Nielsen}}, \ and\ \bibinfo
  {author} {\bibfnamefont {U.~L.}\ \bibnamefont {Andersen}},\ }\href@noop {} {\
  \emph {\bibinfo {title} {A fault-tolerant continuous-variable
  measurement-based quantum computation architecture}}},\ (2021), \bibinfo {note}
  {arxiv:2101.03014}\BibitemShut {NoStop}%
\bibitem [{\citenamefont {Vahlbruch}\ \emph {et~al.}(2008)\citenamefont
  {Vahlbruch}, \citenamefont {Mehmet}, \citenamefont {Chelkowski},
  \citenamefont {Hage}, \citenamefont {Franzen}, \citenamefont {Lastzka},
  \citenamefont {Goßler}, \citenamefont {Danzmann},\ and\ \citenamefont
  {Schnabel}}]{vahlbruch_observation_2008}%
  \BibitemOpen
  \bibfield  {author} {\bibinfo {author} {\bibfnamefont {H.}~\bibnamefont
  {Vahlbruch}}, \bibinfo {author} {\bibfnamefont {M.}~\bibnamefont {Mehmet}},
  \bibinfo {author} {\bibfnamefont {S.}~\bibnamefont {Chelkowski}}, \bibinfo
  {author} {\bibfnamefont {B.}~\bibnamefont {Hage}}, \bibinfo {author}
  {\bibfnamefont {A.}~\bibnamefont {Franzen}}, \bibinfo {author} {\bibfnamefont
  {N.}~\bibnamefont {Lastzka}}, \bibinfo {author} {\bibfnamefont
  {S.}~\bibnamefont {Goßler}}, \bibinfo {author} {\bibfnamefont
  {K.}~\bibnamefont {Danzmann}}, \ and\ \bibinfo {author} {\bibfnamefont
  {R.}~\bibnamefont {Schnabel}},\ }\bibfield  {title} {\enquote {\bibinfo
  {title} {Observation of {Squeezed} {Light} with 10-{dB} {Quantum}-{Noise}
  {Reduction}},}\ }\href {\doibase 10.1103/PhysRevLett.100.033602} {\bibfield
  {journal} {\bibinfo  {journal} {Phys. Rev. Lett.}\ }\textbf {\bibinfo
  {volume} {100}},\ \bibinfo {pages} {033602} (\bibinfo {year} {2008})}.
\bibitem [{\citenamefont {Li}\ \emph {et~al.}(2020)\citenamefont {Li},
  \citenamefont {Li},\ and\ \citenamefont {Agarwal}}]{li_experimental_2020}%
  \BibitemOpen
  \bibfield  {author} {\bibinfo {author} {\bibfnamefont {F.}~\bibnamefont
  {Li}}, \bibinfo {author} {\bibfnamefont {T.}~\bibnamefont {Li}}, \ and\
  \bibinfo {author} {\bibfnamefont {G.~S.}\ \bibnamefont {Agarwal}},\
  }\href@noop {} {\ \emph {\bibinfo {title} {Experimental study of decoherence
  of the two-mode squeezed vacuum state via second harmonic generation}}},\ (2020),\
  \bibinfo {note} {arxiv:2012.11839}\BibitemShut {NoStop}%
\bibitem [{\citenamefont {Ren}\ \emph {et~al.}(2021)}]{ren_topologically_2021}%
  \BibitemOpen
  \bibfield  {author} {\bibinfo {author} {\bibfnamefont {R.}~\bibnamefont {Ren}}, \bibinfo {author} {\bibfnamefont {Y.}~\bibnamefont {Lu}}, \bibinfo {author} {\bibfnamefont {Z.}~\bibnamefont {Jiang}}, \bibinfo {author} {\bibfnamefont {J.}~\bibnamefont {Gao}}, \bibinfo {author} {\bibfnamefont {W.}~\bibnamefont {Zhou}}, \bibinfo {author} {\bibfnamefont {Y.}~\bibnamefont {Wang}}, \bibinfo {author} {\bibfnamefont {Z.}~\bibnamefont {Jiao}}, \bibinfo {author} {\bibfnamefont {X.}~\bibnamefont {Wang}}, \bibinfo {author} {\bibfnamefont {A.}~\bibnamefont {Solntsev}} \ and\
  \bibinfo {author} {\bibfnamefont {X.}\ \bibnamefont {Jin}},\
  }\href@noop {} {\ \emph {\bibinfo {title} {Topologically Protecting Squeezed Light on a Photonic Chip}}},\ (2021),\
  \bibinfo {note} {arxiv:2106.07425}\BibitemShut {NoStop}%
\bibitem [{\citenamefont {Su}\ \emph {et~al.}(1979)\citenamefont {Su},
  \citenamefont {Schrieffer},\ and\ \citenamefont {Heeger}}]{su_solitons_1979}%
  \BibitemOpen
  \bibfield  {author} {\bibinfo {author} {\bibfnamefont {W.~P.}\ \bibnamefont
  {Su}}, \bibinfo {author} {\bibfnamefont {J.~R.}\ \bibnamefont {Schrieffer}},
  \ and\ \bibinfo {author} {\bibfnamefont {A.~J.}\ \bibnamefont {Heeger}},\
  }\bibfield  {title} {\enquote {\bibinfo {title} {Solitons in
  {Polyacetylene}},}\ }\href {\doibase 10.1103/PhysRevLett.42.1698} {\bibfield
  {journal} {\bibinfo  {journal} {Physical Review Letters}\ }\textbf {\bibinfo
  {volume} {42}},\ \bibinfo {pages} {1698} (\bibinfo {year}
  {1979})}\BibitemShut {NoStop}%
\bibitem [{\citenamefont {Heeger}\ \emph {et~al.}(1988)\citenamefont {Heeger},
  \citenamefont {Kivelson}, \citenamefont {Schrieffer},\ and\ \citenamefont
  {Su}}]{heeger_solitons_1988}%
  \BibitemOpen
  \bibfield  {author} {\bibinfo {author} {\bibfnamefont {A.~J.}\ \bibnamefont
  {Heeger}}, \bibinfo {author} {\bibfnamefont {S.}~\bibnamefont {Kivelson}},
  \bibinfo {author} {\bibfnamefont {J.~R.}\ \bibnamefont {Schrieffer}}, \ and\
  \bibinfo {author} {\bibfnamefont {W.~P.}\ \bibnamefont {Su}},\ }\bibfield
  {title} {\enquote {\bibinfo {title} {Solitons in conducting polymers},}\
  }\href {\doibase 10.1103/RevModPhys.60.781} {\bibfield  {journal} {\bibinfo
  {journal} {Reviews of Modern Physics}\ }\textbf {\bibinfo {volume} {60}},\
  \bibinfo {pages} {781} (\bibinfo {year} {1988})}\BibitemShut {NoStop}%
\bibitem [{\citenamefont {Haus}(1991)}]{haus_coupled-mode_1991}%
  \BibitemOpen
  \bibfield  {author} {\bibinfo {author} {\bibfnamefont {H.~A.}~\bibnamefont{Haus}},\ and \bibinfo {author} {\bibfnamefont {W.}~\bibnamefont{Huang}},\ }
  \bibfield  {title} {\enquote {\bibinfo {title} {Coupled-Mode Theory},}\ }\href
  {\doibase 10.1109/5.104225} {\bibfield  {journal} {\bibinfo
  {journal} {Proceedings of the IEEE}\ }\textbf {\bibinfo {volume} {79}},\
  \bibinfo {pages} {1505} (\bibinfo {year} {1991})}\BibitemShut {NoStop}%
\bibitem [{\citenamefont {Asboth}(2016)}]{asboth_short_2016}%
  \BibitemOpen
  \bibfield  {author} {\bibinfo {author} {\bibfnamefont {J.~K.}~\bibnamefont{Asb\'{o}th}},\ \bibinfo {author} {\bibfnamefont {L.}~\bibnamefont{Oroszl\'{a}ny}},\ and \bibinfo {author} {\bibfnamefont {A.}~\bibnamefont{P\'{a}lyi}},\ }
  \bibfield  {title} {\enquote {\bibinfo {title} {A Short Course on Topological Insulators},}\ } Lecture Notes in Physics, Vol.~919 (Springer International Publishing, 1991) \BibitemShut {NoStop}%
\bibitem [{\citenamefont {Rojas-Rojas}(2019)}]{rojas-rojas_manipulation_2019}%
  \BibitemOpen
  \bibfield  {author} {\bibinfo {author} {\bibfnamefont {S.}~\bibnamefont{Rojas-Rojas}},\ \bibinfo {author} {\bibfnamefont {E.}~\bibnamefont{Barriga}},\ \bibinfo {author} {\bibfnamefont {C.}~\bibnamefont{Muñoz}},\ \bibinfo {author} {\bibfnamefont {P.}~\bibnamefont{Solano}},\ and \bibinfo {author} {\bibfnamefont {C.}~\bibnamefont{Hermann-Avigliano}},\ }
  \bibfield  {title} {\enquote {\bibinfo {title} {Manipulation of multimode squeezing in a coupled waveguide array},}\ }\href
  {\doibase 10.1103/PhysRevA.100.023841} {\bibfield  {journal} {\bibinfo
  {journal} {Phys. Rev. A}\ }\textbf {\bibinfo {volume} {100}},\
  \bibinfo {pages} {023841} (\bibinfo {year} {2019})}\BibitemShut {NoStop}%
\bibitem [{\citenamefont {Marino}(2006)}]{marino_experimental_2006}%
  \BibitemOpen
  \bibfield  {author} {\bibinfo {author} {\bibfnamefont {A.}~\bibnamefont{Marino}},\ }
  \bibfield  {title} {\enquote {\bibinfo {title} {Experimental studies of two-mode squeezed states in rubidium vapor},}\ } Ph.D. thesis, The Institute of Optics, The College School of Engineering and Applied Science, Univerity of Rochester, Rochester, New York (2006)\BibitemShut {NoStop}%
\bibitem [{\citenamefont {Simon}(2000)}]{simon_peres-horodecki_2000}%
  \BibitemOpen
  \bibfield  {author} {\bibinfo {author} {\bibfnamefont {R.}~\bibnamefont
  {Simon}},\ }\bibfield  {title} {\enquote {\bibinfo {title} {Peres-{Horodecki}
  {Separability} {Criterion} for {Continuous} {Variable} {Systems}},}\ }\href
  {\doibase 10.1103/PhysRevLett.84.2726} {\bibfield  {journal} {\bibinfo
  {journal} {Physical Review Letters}\ }\textbf {\bibinfo {volume} {84}},\
  \bibinfo {pages} {2726} (\bibinfo {year} {2000})}\BibitemShut {NoStop}%
\end{thebibliography}
%

\onecolumn\newpage
\appendix

\section{Numerical simulation of propagation of squeezed states}
\label{appendix:A}

In this section we detail how the propagation of squeezed states is realized by numerical simulation. Our method relies on the fact that squeezed vacuum states are Gaussian with zero mean value of the quadratures, and are therefore fully described by their correlation matrix, which can be obtained from matrices $N$ and $M$ defined by their matrix elements $N_{n, m} = \langle a_n^\dag a_m \rangle$ and $M_{n, m} = \langle a_n a_m \rangle$. Thus the problem reduces to simulating the evolution of both matrices.

Let $\hat{H}$ be the hamiltonian operator of the system, so the annihilation operators $a_n$ satisfy the Heisenberg equation 

\begin{equation}
    -i \frac{da_n}{dz} = \left[\hat{H}, a_n\right]
\end{equation}

The solution to this equation is $a_n(z) = \hat{U}^\dag(z) a_n(0) \hat{U}(z)$, with $\hat{U}(z) = \exp(-i\hat{H}z)$ the $z$ evolution operator. We represent the hamiltonian operator in matrix form, under the basis of the annihilation operators of the waveguides $a_n$, as

\begin{equation}
    \hat{H} = \sum_{n, m} a_n^\dag H_{n, m} a_m \text{ .}
    \label{eq:hamiltonianmatrix}
\end{equation}
Let $Q$ be the matrix containing the eigenvectors of $H$ in its columns, so $Q^\dag H Q = \tilde{H}$, with $\tilde{H}$ the diagonal form of $H$, containing the propagation constant of each eigenmode. The $z$ evolution operator is represented in matrix form by $U = Q \exp(-i\tilde{H}z)Q^\dag$, and the $z$ evolution of the waveguide operators is

\begin{equation}
    a_n(z) = \sum_m U_{n, m}(z) a_m(0)
\end{equation}

We finally obtain the evolution of the $N$ and $M$ matrices, described by

\begin{align}
    N_{n, m}(z) = \sum_{k, l} U_{n, k}^*(z) U_{m, l}(z) N_{k, l}(0) \text{ ,} \\
    M_{n, m}(z) = \sum_{k, l} U_{n, k}(z) U_{m, l}(z) M_{k, l}(0) \text{ .}
\end{align}

\section{Extended results of section \ref{subsection:1mB}}
\label{appendix:B}

In this section we present a more complete report of the results discussed in section \ref{subsection:1mB} of the main text, that could help the reader follow the corresponding discussions. For all figures we have used $\alpha=0.3$, and $\xi=0.9$.

\begin{figure}[h!]
    \centering
    \includegraphics[width=0.7\textwidth]{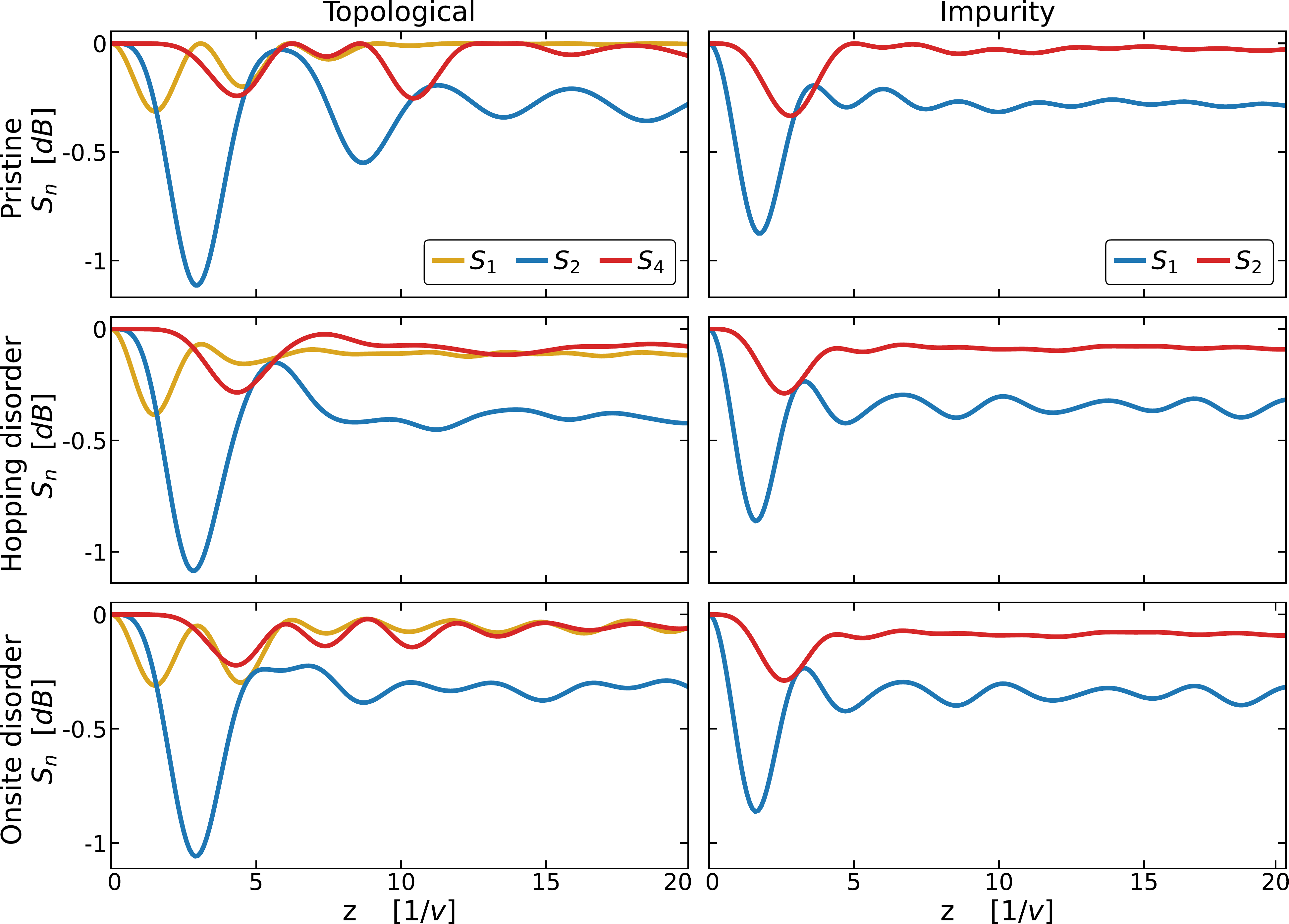}
    \caption{One-mode squeezing along the propagation axis, for $\alpha=0.3$ and $\xi=0.9$. Top, middle and bottom row shows one-mode squeezing for pristine, hopping disordered ($d=0.6v$), and onsite disordered ($d=0.6v$) lattices respectively, while left and right columns refer to the topological and impurity states respectively. The site at which the measurement is taken is indicated in the legend, which is valid for the entire column.}
    \label{fig:1m_1msqueezing}
\end{figure}

In Fig.~\ref{fig:1m_1msqueezing} we show one-mode squeezing along the propagation axis at waveguides $1$, $2$ and $4$ of the topological lattice, and waveguides $1$ and $2$ of the impurity one. These results allow to envision the spatial distribution of one-mode squeezing across the lattices.

\begin{figure}[h!]
    \centering
    \includegraphics[width=0.7\textwidth]{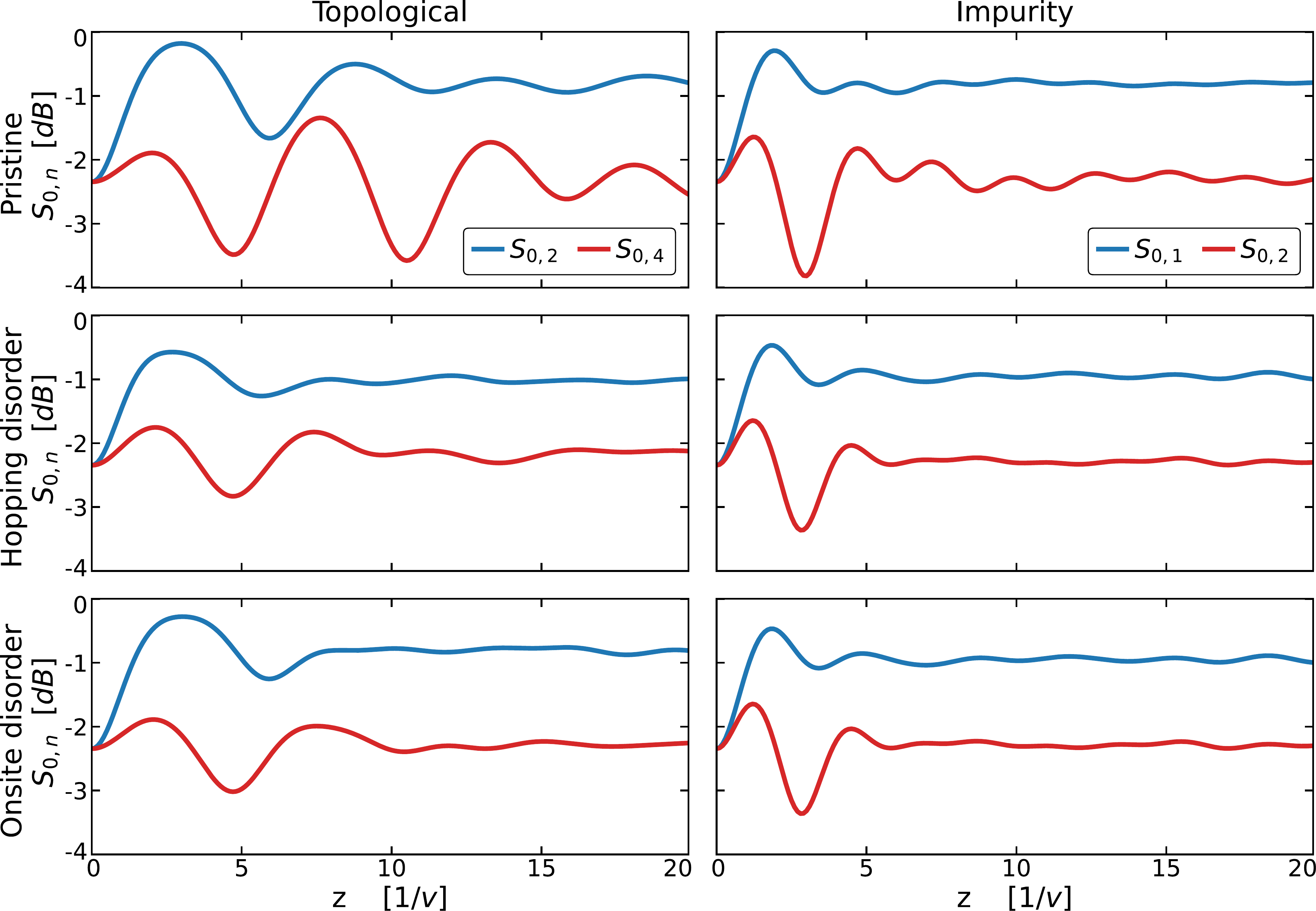}
    \caption{Two-mode squeezing along the propagation axis, for $\alpha=0.3$ and $\xi=0.9$. Top, middle and bottom row shows two-mode squeezing for pristine, hopping disorder ($d=0.6v$), and onsite disordered ($d=0.6v$) lattices respectively, while left and right columns refer to the topological and impurity states respectively. The site at which the measurement is taken is indicated in the legend, which is valid for the entire column.}
    \label{fig:1m_2msqueezing}
\end{figure}

In Fig.~\ref{fig:1m_2msqueezing} we show two-mode squeezing along the propagation axis, measured between the edge waveguide and the first two waveguides that take part in the localized eigenmode, namely, waveguides 2 and 4 of the topological lattice, and waveguides 1 and 2 of the impurity one. This shows the effects of the phase relation between the modes on two-mode squeezing.

\newpage
\section{On the statistical fluctuations for Fig.~\ref{fig:1m_squeezing}~and~\ref{fig:2m_squeezing}}

\begin{figure}[h!]
    \centering
    \includegraphics[width=0.9\textwidth]{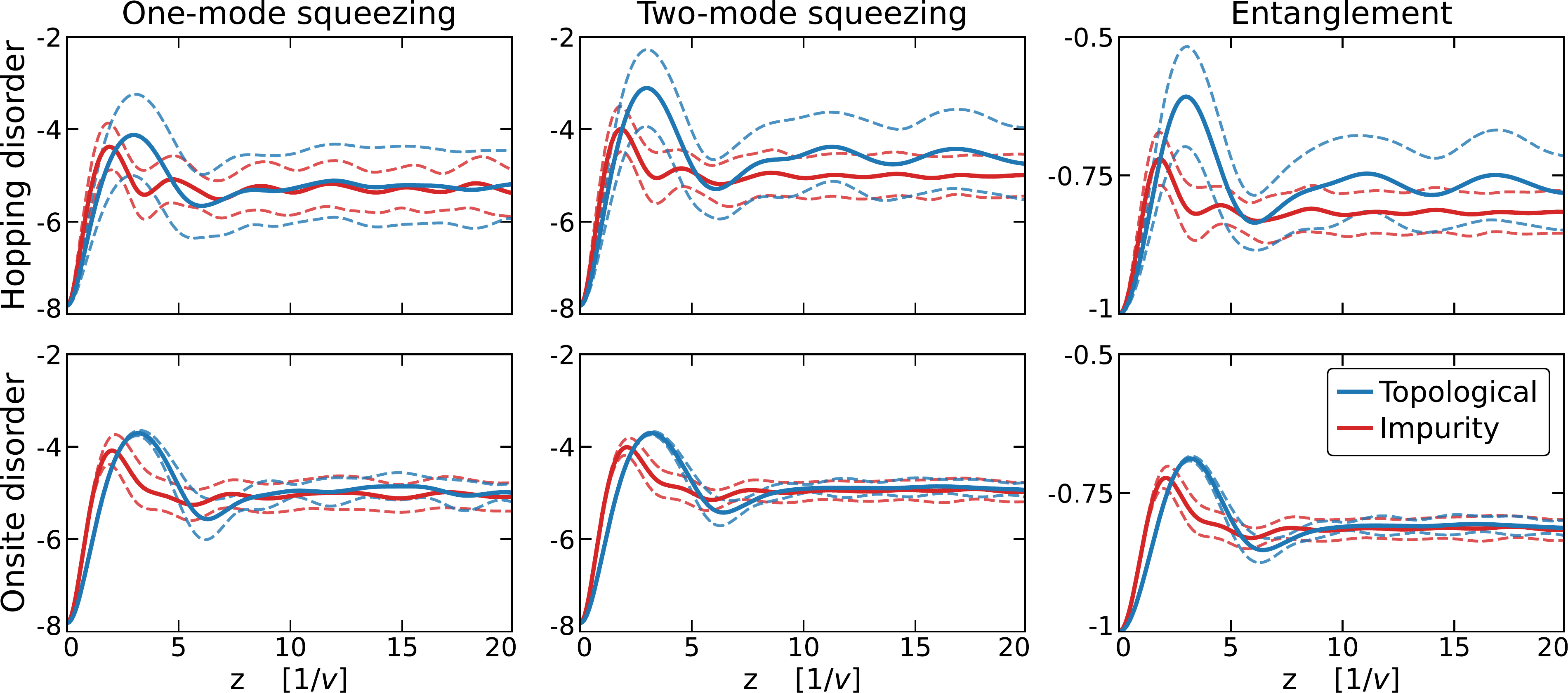}
    \caption{Maximal one-mode squeezing corresponding to the data in Fig.~\ref{fig:1m_squeezing}, and maximal two-mode squeezing and entanglement corresponding to the data in Fig.~\ref{fig:2m_squeezing}. Statistical average values are represented by solid lines, while dashed lines represent the confidence interval associated with the standard deviation, with blue (red) lines referring to the topological (impurity state).}
    \label{fig:statisticalfluctuations}
\end{figure}

In this section we refer to the statistical fluctuations for the data presented in Fig.~\ref{fig:1m_squeezing} and \ref{fig:2m_squeezing} of the main text. In Fig.~\ref{fig:statisticalfluctuations} we show maximal one-mode squeezing corresponding to the data in Fig.~\ref{fig:1m_squeezing}, and maximal two-mode squeezing and entanglement corresponding to the data in Fig.~\ref{fig:2m_squeezing}, along with the confidence intervals of the corresponding quantities. We observe that after the transient response of the system when the initial state is injected at the edge waveguide, the curves of the topological and impurity state take approximately equal values, falling within the confidence interval of each other. We also observe that hopping disorder induced fluctuations are considerably larger than those generated by onsite disorder in both systems. Furthermore, fluctuations in the hopping disordered topological state are larger than those of the impurity one. However, the magnitude of the fluctuations is consistently lower than the magnitude of the physical quantities themselves in all cases. We have checked as well that increasing the number of random realizations yields narrower confidence intervals, and that higher disorder values yield larger confidence intervals, in tune with th behavior observed in Fig.~\ref{fig:ideal}.

\section{Movie 1: Caption}

Evolution of the average Wigner function at the zero-th waveguide when the lattices are initially excited with single-mode squeezed state, as studied in section \ref{subsection:1mB}, for $\alpha=0.3$ and $\xi=0.9$. Top, middle and bottom row refer to the pristine, hopping disordered ($d=0.6v$), and onsite disordered ($d=0.6v$) lattices respectively, while left and right columns refer to the topological, and impurity induced (topologically trivial) states respectively. The bar at the bottom indicates the distance along the propagation axis $z$. When chiral symmetry is preserved (pristine and hoping disorder) the maximally squeezed quadrature of the topological state is always $X_1$ as its propagation constant is that of the bare waveguides, therefore the ellipse remains oriented as the initial state. In contrast, when the symmetry is broken (onsite disorder), the maximally squeezed quadrature differs between different random realizations, thus no squeezing is obtained when averaging over them. Because the impurity induced state does not present any symmetry protected properties, upon any type of disorder the maximally squeezed quadrature fluctuates between the different random realizations and no average squeezing is measured in either case.

\end{document}